\documentclass[a4paper,12pt]{article}

\usepackage{amssymb,amsmath,mathbbol,mathrsfs}
\usepackage[usenames,dvipsnames]{color}
\usepackage{hyperref}
\usepackage{xcolor}
\usepackage{stmaryrd}
\usepackage{authblk}
\usepackage{framed}
\usepackage{empheq}
\usepackage{slashed}
\usepackage{hyphenat}
\usepackage{cite}
\usepackage{apacite}
\usepackage{natbib}
\usepackage[normalem]{ulem} 
\usepackage{soul}

\usepackage{dsfont}
\usepackage{chngcntr}
\usepackage{enumerate}

\usepackage{marginnote}    

\usepackage[left=.8in,right=.8in,top=.8in,bottom=.8in]{geometry}                  

\usepackage{sectsty}
\allsectionsfont{\sffamily\mdseries\upshape} 
\usepackage{tocloft}

\makeatletter
\renewenvironment{abstract}{%
    \if@twocolumn
      \section*{\abstractname}%
    \else 
      \begin{center}%
        {\bfseries\sffamily\abstractname\vspace{\z@}}
      \end{center}%
      \quotation
    \fi}
    {\if@twocolumn\else\endquotation\fi}
\makeatother

\numberwithin{equation}{section}
5

\hypersetup{
	colorlinks=true,         
	linkcolor=MidnightBlue,          
	citecolor=BrickRed,        
	urlcolor=MidnightBlue            
}

\newcommand\mathC{\mkern1mu\raise2.2pt\hbox{$\scriptscriptstyle|$}
        {\mkern-7mu\rm C}}              

\newcommand{\R}{{\mathbb{R}}}

\newcommand{\be}{\begin{equation}}
\newcommand{\ee}{\end{equation}}

\newcommand{\fg}{{\mathfrak{g}}}

\newcommand{\fLie}{{\mathbb{L}}}
\newcommand{\fI}{{\mathbb{i}}}
\newcommand{\F}{{{\Phi}}}
\renewcommand{\d}{{\mathrm{d}}}
\newcommand{\D}{{\mathrm{D}}}

\newcommand{\G}{{\mathcal{G}}}
\newcommand{\Ad}{{\mathrm{Ad}}}
\newcommand{\pp}{{\partial}}

\newcommand{\rr}{{\mathbf{r}}}
\newcommand{\SO}{{\mathrm{SO}}}

\newcommand{\cc}{{\mathbf{c}}}

\newcommand{\fF}{{\mathbb{F}}}

\newcommand{\fG}{{\mathrm{Lie}(\G)}}

\newcommand{\dd}{{\mathbb{d}}}

\renewcommand{\hat}{\widehat}

\newcommand{\RR}{\mathds{R}} 

\newcommand{\cint}{{\int\kern-.87em{<}}}
\newcommand{\sint}{{\int\kern-.75em{\sim}}}
\newcommand{\fint}{{\int\kern-1.00em{\int}}}

\newcommand{\bb}{\mathbb}

\renewcommand{\#}{\sharp}

\let\oldmarginpar\marginpar
\renewcommand\marginpar[1]{\oldmarginpar{\color{red}\raggedright\footnotesize #1}}

\newcommand{\jb}{\color{ForestGreen}}

\newcommand{\old}{\color{red}}

\begin{document}

\title{The Hole Argument and Beyond:\\
 Part II: Treating Non-isomorphic Spacetimes}

\author{Henrique Gomes and Jeremy Butterfield\footnote{\href{mailto:gomes.ha@gmail.com}{gomes.ha@gmail.com, jb56@cam.ac.uk}} \\\it  Oriel College, University of Oxford; Trinity College, University of Cambridge; }

\maketitle
\vspace{-1cm}
\begin{abstract}
 In this two-part paper we review, and then develop, the assessment of the hole argument for general relativity. 

The review (in Part I) discussed how to compare points in isomorphic spacetimes, i.e. models of the theory. This second Part proposes a framework for making comparisons of {\em non}-isomorphic spacetimes. It combines two ideas we discussed in Part I---the philosophical idea of {\em counterparts}, and the idea of threading points between spacetimes other than by isomorphism---with the mathematics of fibre bundles. 

We first recall the ideas from Part I (Section 1). Then in Section 2 and an Appendix, we define a fibre bundle whose fibres are isomorphic copies of a given spacetime or model, and discuss connections on this fibre bundle. This material proceeds on analogy with field-space formulations of gauge theories. Finally, in Section 3, we show how this fibre bundle gives natural expressions of the philosophical ideas of counterparts, and of threading.

 \end{abstract}
  
\newpage

\tableofcontents 

\newpage

 \section{Introduction}\label{intro}
  
 In Part I of this paper, we reviewed, mostly from a philosophical perspective, the hole argument in general relativity. It is about comparing isomorphic spacetimes (i.e. solutions of general relativity), and especially about how to ``identify’’ a point in one spacetime with a point in another isomorphic one. 

Here in Part II, we break new ground. We address the question of how one should compare {\em non}-isomorphic models: in particular, how to compare points between the two models.  We will propose a framework for such comparisons, that combines the mathematics of fibre bundles with two ideas we developed in Part I:\\
\indent \indent (i): the philosophical idea of {\em counterparts}: which is a way of treating the identity of objects---any objects, not only points---in different possible situations; (cf. Section 2.2.3 of Part I); and :\\
\indent \indent (ii) the idea of {\em threading points}: which is our jargon for a point in one spacetime corresponding to (being ``identified” with) a point in another spacetime {\em other than by} their being, respectively, the argument and value of (a diffeomorphism that is) an isomorphism between the spacetimes; (cf. Section 3.2 of Part I).

Our proposed framework proceeds by analogy with field-space formulations of gauge theories. It postulates a fibre bundle whose fibres are isomorphic copies of a given model, i.e. a given Lorentzian manifold endowed with matter fields obeying the Einstein equations. So given a model based on a manifold $M$, Diff($M$) is to be the structure group of the bundle: any element of Diff($M$) drags the given model into coincidence (and so isomorphism) with another element of the fibre. So both the base-manifold (the set of orbits) of the bundle, and its structure group, are infinite-dimensional; and our exposition will thus need to address subtleties about infinite-dimensional bundles.\footnote{\label{maybesmaller} Although we will throughout this paper talk of Diff($M$) as the structure group, almost all of our discussion would hold good  for a smaller group. And since we think of the elements of a fibre as physically equivalent, such a smaller group would be motivated by the standard  practice in general relativity of regarding some isomorphisms as generating new possibilities. The   main example is isomorphisms that do not preserve structure at asymptotic spatial infinity (but correspond to e.g. a time translation there). For discussion and references, cf. footnote 24 of Part I, and footnote \ref{Belot2} below.  In any case: even if we replace Diff($M$) by some subgroup whose elements also preserve some desired asymptotic structure, both the bundle and this smaller structure group, would still be infinite-dimensional.} \\

 The paper is organised as follows. In Section \ref{sec:4fibre}, we expound the fibre bundle of spacetimes. This exposition will use a natural analogy with ideas and results in classical gauge theories (Yang-Mills theories, both abelian and non-abelian)---for which of course the appropriateness of fibre bundle ideas is universally accepted. Though our bundle is infinite-dimensional, we shall see that much carries over from the familiar finite-dimensional case. We will first describe the bundle (Section \ref{subsec:41bdle}); and then describe connections on it (Section \ref{subsec:42connn};  and in the Appendix, Sections \ref{sec:SdW_YM} and \ref{sec:SdW_GR}). For the main tool for comparing elements of distinct fibres of any bundle is a connection: so in order to compare non-isomorphic spacetimes in our bundle, we need to consider connections.  Then in Section \ref{sec:5cpart} , we explain how Section \ref{sec:4fibre}'s material relates to the idea of {\em threading}, and also gives a natural and technically precise illustration of {\em counterparts}.  

Two final orienting remarks. (1): Broadly speaking, the ingredients of Section \ref{sec:4fibre} come from physics; and Section \ref{sec:5cpart} adds ingredients from philosophy. Unsurprisingly, this makes Section \ref{sec:4fibre} more technical: especially since our fibre bundle of spacetimes is infinite-dimensional, while the fibre bundles familiar from gauge theory are finite-dimensional,  and have space or spacetime as the base manifold.  On the other hand, the technicalities in Section \ref{sec:4fibre} are already in the literature; and so we can be brief, with details in the Appendix. Besides, these technicalities will not be needed for Section \ref{sec:5cpart}'s philosophical ``punch-line": that our bundle of spacetimes illustrates counterpart theory, with neat mathematical properties. Indeed, that punch-line will use little more than the idea of a section of the bundle of spacetimes.

(2):  We should note at the outset how this paper uses the word `point’ in two different ways; (though the meaning will always be clear from the context). For the fibre bundles in the familiar formulations of gauge theories, the base-manifold is spacetime or space: whose elements are thus points of exactly the kind at issue in the hole argument and in our Part I. And a point in, i.e. element of, the bundle itself is an assignment of field-values to a spacetime (or spatial) point. On the other hand, for the infinite-dimensional bundles of this paper (both those for general relativity, and those for gauge theories), a point in, i.e. element of, the bundle is a field-configuration across all of spacetime (space). And accordingly, a point in the base-manifold gives the physical information (what Part I called `qualitative profile') shared by all the field-configurations in the fibre above the point---the 
 ``gauge  redundancy" having been quotiented out.

So in short: In Part I and in the hole argument, one naturally reads `point’ as a spacetime (or spatial) point, and in that context, mention of a point in a fibre bundle suggests ``at most” an assignment of field-values to a spacetime (or spatial) point (and so a familiar finite-dimensional bundle). But in this paper, the points in the fibre bundles that we will be concerned with are field-configurations across all of spacetime (space). But no matter: which meaning of `point' is intended will always be clear from the context.

 \section{A fibre bundle of models}\label{sec:4fibre}

In this Section, we will proceed in two stages. In Section \ref{subsec:41bdle}, we will introduce the infinite-dimensional fibre bundle of Lorentzian manifolds, with
 the diffeomorphism group as its structure group. In classical gauge theories,  the analogue of this fibre bundle is {\em field-space}, with the group of gauge transformations as the structure group. Then in Section \ref{subsec:42connn} and the Appendix, we will give some details about connections on these bundles. For a connection is the main tool for comparing elements of distinct fibres, in both finite and infinite-dimensional bundles; (though as just  stressed in (2) of Section \ref{intro}, the elements i.e. `points' being compared  are very different).

\subsection{The bundle of spacetimes}\label{subsec:41bdle} 
The main idea of this Subsection is that given a spacetime $(M,g,T)$, the set of all  models isomorphic to it, that are built by drag-along (in the manner of the hole argument) by all the diffeomorphisms $d$ on $M$ is   like
 the fibre of a principal fibre bundle, whose structure group is Diff($M$), the group of diffeomorphisms of $M$. We will first: (1) state this idea more precisely, and state the analogy with gauge theory's idea of {\em field-space}; and then (2) give some technical qualifications, which apply to both general relativity and gauge theory. \\

\noindent (1): {\em The idea, and the analogy with gauge theory:---}  First, we fix the spacetime manifold $M$; although general relativity has models using non-diffeomorphic manifolds, all that follows uses a single $M$.

 For simplicity, let us also  set aside $T$, so that we take general relativity {\em in vacuo}.  Then the class of all Lorentzian manifolds $(M,g)$ built on our fixed $M$ is obviously partitioned into isometry classes by Diff($M$). Our task now is to describe how this class has further structure, making it an infinite-dimensional bundle, of which each fibre is an orbit of Diff($M$), i.e. an isometry class. 

The situation is, in spirit, identical to how in gauge theory one can treat {\em field-space}, i.e. the space of field configurations, as an infinite-dimensional bundle, of which each fibre is an orbit of the group of gauge transformations. Note that here `field configuration' means
 an assignment of values to the fields, including gauge fields, throughout all space, or all spacetime.  
 So field-space is infinite-dimensional, and partitioned into gauge-equivalence classes by the gauge transformations. Since as usual we allow the gauge transformations to be local, i.e. to depend on the spatial or spacetime point, the group of gauge transformations is obviously an infinite-dimensional group. So here, the parallel task  is to describe how  field-space has further structure, making it an infinite-dimensional bundle, of which each fibre is an orbit of the group of gauge transformations.

Since the main idea is the same in both general relativity and gauge theories, we will adopt a common language and notation.\footnote{The exposition in this Subsection and in the next is drawn from \cite{Samediff_0, Samediff_1a, Samediff_1b}. Further references  are given below.} The theory gives us a space $\Phi$ of models $\varphi$. For general relativity  {\em in vacuo}, $\varphi$ is a Lorentzian manifold $(M,g)$, while for gauge theory $\varphi$ is a field configuration. A group $\cal G$ acts on $\Phi$, and this action respects the manifold structure of $\Phi$. In our examples, $\cal G$ is:   for general relativity, the set of diffeomorphisms of $M$;  and for gauge theory, the set of 
gauge transformations. We denote the action by: ${\cal G} \times {\Phi} \ni (g,\varphi) \mapsto \varphi^g \in \Phi$. So in particular, for each $g$, the map on $\Phi$ thereby defined, i.e. $\varphi \mapsto \varphi^g$, is a diffeomorphism of $\Phi$.

We write an orbit of this action with the usual square-bracket notation for equivalence classes: $ [\varphi]$, where $\pi:\F\rightarrow [\F]$ denotes the surjective projection onto the set of equivalence classes,  which we write as  $[\F]$. 
And the set of elements in $\F$ that are equivalent to $\varphi$ is also denoted by   $O_{\varphi}:=\pi^{-1}([\varphi])$, to stress that it is an orbit.
So the equivalence class  gives the ``physical information” shared by all the elements of the orbit. And once a bundle structure is established, then, as mentioned in the preamble to this Section: an element of the base-manifold (which in some definitions of a bundle, is defined to be the set of equivalence classes) gives the physical information shared by all the points in the fibre above the element. In the jargon of gauge theory: an element of the base-manifold is the {\em physical state}: the  gauge redundancy having been quotiented out. So for our example of general relativity, i.e. the bundle of Lorentzian manifolds $(M,g)$ on a given manifold $M$, an element of the base-manifold is the qualitative profile of metric and matter field values, which quotients out ``which spacetime point is which'' within $M$.\footnote{\label{Belot2}{Here, we should qualify our statement above that Diff($M$) is the structure group. For recall footnote \ref{maybesmaller} and footnote 24   of Part I: for spacetimes that are asymptotically flat at infinity, it is standard practice to treat some isomorphisms (for vacuum spacetimes: some isometries) as generating from a given model a new physical possibility, that differs from the given model by  e.g. 
a time-translation at spatial infinity. (For a masterly philosophical discussion, cf. \cite{Belot50}, especially Sections 4.3 and 4.4, pages 964-970.) To allow for this, while maintaining our interpretation that elements of a given fibre are all physically equivalent, we need to allow that the structure group of the bundle of spacetimes is smaller than Diff($M$), i.e. its elements fix the structure at spatial infinity. But this qualification does not affect anything in the rest of this paper. Besides, to stress the analogy with field-space formulations of gauge theory, we are anyway writing the structure group as $\cal G$.}}

Besides, general relativity and gauge theories are not the only physically relevant examples of an infinite-dimensional space $\Phi$ of models $\varphi$, on which a group $\cal G$ acts. Another salient example is the Riemannian analogue of our Lorentzian bundle, i.e. the bundle of Riemannian manifolds $(M,g)$ on a given manifold $M$: in other words, the configuration space of Riemannian metrics $g$ on $M$. It is often written as Riem($M$). Similarly to the Lorentzian case, a fibre of Riem($M$) is built from a given Riemannian manifold (`configuration') $(M,g)$ by drag-along, in the manner of the hole argument, by any of the diffeomorphisms $d$ on $M$. So here:  an equivalence class under the action of the group quotients out ``which {\em spatial} point is which''. (We will return to this example in footnote \ref{ftnt:stab} and at the end of  Section 4.2.)

 Now we think of the set of   equivalence classes, $[\Phi]$, as itself a manifold. Then we consider smooth maps $\sigma$ from $[\Phi]$ to $\Phi$ that send each argument $[\varphi]$ to an element of  the corresponding orbit, i.e. $\sigma([\varphi]) \in \mathcal{O}_\varphi$. In short, $\sigma$ is a smooth choice of a {\em representative} of each orbit. When the manifold $\F$ has a principal fiber bundle structure, such a map is a {\em section} of $\Phi$; the word is also used for the map's range, i.e. the embedded sub-manifold of $\Phi$ that crosses each orbit only once. In the context of a principal bundle, given some open subset $\mathcal{U}\subset \F$, a section would allow us to write a local product structure, i.e. a diffeomorphism between  $\pi^{-1}([\mathcal{U}])$ and $[\mathcal{U}]\times \G$.

So to summarise: the main idea is that the space of models $\F$ is very similar to a principal fiber bundle, with $\G$ (respectively: diffeomorphisms and vertical automorphisms) as its structure group. But there are important differences between these infinite-dimensional sets and the finite-dimensional principal fiber bundles seen in familiar formulations of gauge theories. In the finite-dimensional case, it suffices that the action of a group $G$ on a given manifold $P$ be free and proper for that manifold to have a principal ${G}$-bundle structure. (This structure is usually written as ${G}\hookrightarrow P\rightarrow P/{G}$, where $P/{G}$ is the base-manifold.)  But in the infinite-dimensional case, these properties of the group action are not enough to guarantee the necessary fibered, or local product, structure. This is taken up in (2). \\

\noindent (2): {\em Qualifications:---}  Indeed, there are three qualifications to be made here.

The first is that there are symmetric $\varphi$, i.e. $\varphi$ that are fixed by some $g \in {\cal G}$,  i.e.  $\varphi^g = \varphi$   (where of course $g$ is not the identity element). Such a $\varphi$ is called {\em reducible}, and the $g$ that fix it are called \emph{stabilisers}.  
It is easy to show that all elements of a reducible state’s orbit are reducible (with stabilisers related by  conjugacy). 
  This  means that the orbits are of various ``sizes”, and are not all isomorphic to the structure group. So $[\Phi]$ is in fact, not a single manifold, but a union of manifolds, one for each ``size” of orbit.  

The manifolds $\Phi$ 
  with a group action  that we discuss here are thus decomposed into different parts, 
   with different sizes of orbits, called {\em strata}. That is,  $\Phi$  is stratified into---is a disjoint union of---orbits of states that possess either more or fewer stabilisers: with the orbits  whose elements have more stabilisers  (i.e. stabiliser groups of larger dimension)  being at the boundary of the orbits of states with fewer stabilisers;   (and with the `bulk' stratum being the manifold consisting of generic fields $\varphi$ that have only the identity element of $\cal G$ as a stabiliser). For each stratum, we can find a section and form a product structure as in the standard picture of the principal bundle.   The notion that generalises sections to cut across  such sets of orbits is called \emph{a slice}, and there are several existence theorems for the case of gauge theories and metrics.\footnote{
Given $\varphi\in\F$, a slice at $\varphi$ is a submanifold  $\Sigma$ of $\F$ containing $\varphi$ such that, if we denote by $\mathsf{Aut}_\varphi\subset \G$ the automorphism, or stabiliser, group of $\varphi$, the following conditions are met: \\
1. $\tilde g\in \mathsf{Aut}_\varphi\Rightarrow \Sigma^{\tilde g}=\Sigma$;\\
2. $g\notin \mathsf{Aut}_\varphi\Rightarrow \Sigma^g\cap \Sigma=\emptyset$;\\
3. There exists a local cross section \emph{of the group} $\tau:Q\subset \G/\mathsf{Aut}_\varphi\rightarrow\G$ where $Q$ is an open neighborhood of the
identity, such that 
\begin{align} \Psi: Q\times \Sigma \rightarrow& \mathcal{U}_\varphi\\
(g,\varphi')\mapsto& \varphi'^{\tau(g)},\end{align}
 where $\mathcal{U}_\varphi$ is an open neighborhood of $\varphi\in\F$, is a diffeomorphism.\\
 Here, we retain the product structure, but it depends on some choice of embedding of (a neighborhood of the identity of) the quotient of the group by the group of automorphisms into the group.}

The second  qualification is that even if we restrict attention to the generic, i.e. not reducible, configurations (for non-abelian field theories)—one can have at most a local product structure: no section is global. This is known as the Gribov obstruction; see \cite{Gribov:1977wm, Singer:1978dk}.  (More precisely: though the name ‘Gribov’ is associated with the infinite-dimensional case, the point here---that a bundle has a local product structure (local trivializations) but not necessarily a global one---applies of course to finite-dimensional bundles.)

So much by way of stating the first two qualifications. The third qualification touches on the topic of sections of the bundle $\Phi$: this will look ahead to Section \ref{subsec:42connn}'s discussion of connections.

 Recall that a section is essentially an embedded submanifold of $\F$ that intersects each orbit exactly once. Since it is natural to define a submanifold of any given manifold $N$ as the level-surface of a real-valued function on $N$, we consider defining a section in this way. More precisely: we aim to define a co-dimension one surface $\Sigma\subset N$,  as $\mathcal{F}^{-1}(c)$, for $c\in \RR$, and $\mathcal{F}$ a smooth and regular function, i.e.  $\mathcal{F}:N\rightarrow \RR$ such that $\d \mathcal{F}\neq 0$. But in the infinite-dimensional case, both the dimension and the co-dimension of a regular value surface can be infinite, and it becomes trickier to construct such a section.

 Since these are infinite-dimensional spaces, they require a more comprehensive definition of manifold. But just like the finite-dimensional manifold is modeled on $\RR^n$, the more general notion of a Frech\'et manifold is modeled on a complete, metrizable, locally convex topological vector space: $\F$ here is always a Frech\'et manifold and $\G$ is a Frech\'et group. These notions allow us to import many of the ideas of differential geometry:  in particular, the idea  of decomposing the total tangent space at a point in a bundle into a direct sum of the tangent space to the orbit, and a complement.  Thus it has been shown using different techniques and at different levels of mathematical rigour (\cite{Ebin, Palais, Mitter:1979un, isenberg1982slice, kondracki1983, YangMillsSlice, Slice_diez}) that both the gauge theory configuration space $\Phi$, and the configuration space of Riemannian metrics on a fixed manifold $M$, have slices and thus admit a local product structure.\footnote{Roughly speaking, one starts by endowing $\F$ with some  $\G$-invariant (super)metric, and then finds the orthogonal complement to the orbits $[\varphi]$, with respect to this supermetric. But here the intersection of the orbit with its orthogonal complement   might not   vanish, as it does in the finite-dimensional case;   and the tangent spaces to the orthogonal section and to the orbit might not  sum to the total tangent space; and they might not be closed under the Frech\'et space topology (based on the compact-open topology). Nonetheless, in the cases at hand, that intersection is given by the kernel of an elliptic operator;  and the tangent space to the  orbits is everywhere a closed subspace of the tangent space to $\F$. Ultimately, this follows from the facts  that (i) the map that embeds $\G$ into $\F$ is an `injective' differential operator (or, more correctly, an operator with injective symbol), and (ii) one can define a canonical, invariant metric on $\F$, that admits a Levi-Civita connection and thus a notion of exponential map.  Thus one can use the Fredholm Alternative  (see \cite[Sec. 5.3 and 5.9]{Trudinger} to show:  that that intersection is at most finite-dimensional, and generically is zero;  that the generic orbit has the `splitting' property, i.e. that the total tangent space decomposes into a direct sum of the tangent space to the orbit and its orthogonal complement; and that we can extend the directions transverse to the orbit, so as to construct a small patch that intersects the neighbouring orbits only once. Unlike the field space of gauge potentials, the space of metrics is not a vector or even an affine space, and so we cannot just linearly extend the directions normal to the orbit. And so we extend   the normal directions by  using the Riemann normal exponential map with respect to the supermetric (cf. \cite{Gil-Medrano}), and thereby conclude that, for a sufficiently small radius, the resulting submanifold is transverse to the neighbouring orbits,   has no caustics,   and is suitably equivariant with respect to the group action (since the exponential map commutes with the diffeomorphisms). Finally, to show that this `section' is not only transverse to the orbits, but indeed that it intersects neighbouring orbits only once, the orbits must be embedded manifolds, and not just local immersions: this is guaranteed if the group action is proper; which it is in the cases of interest. This is guaranteed when the underlying manifold $M$ is compact. When it is not, the fact that the group action from $\G\times \F$ to $\F$ is jointly continuous on $\G$ and $ \F$ in its domain can also be used, \cite[Prop. 130]{Ebin_phd}.
 
 Unfortunately, the space of Lorentzian metrics is {\em not} known to have such a structure. It has only been shown for the space of Einstein metrics, i.e. solutions of the Einstein field equations, that admit a constant-mean-curvature (CMC) foliation and that have a compact spatial Cauchy surface (\cite{isenberg1982slice}).
  The reason is that the CMC-foliability implies a lot of control on the diffeomorphisms orthogonal to the leaves.\label{ftnt:stab}  }

This discussion leads to the topic of {\em connections}, which we take up in the next Subsection.

\subsection{Connections on these bundles}\label{subsec:42connn}

In this Subsection, we note some properties that a {\em connection} on a principal fibre bundle can have. The basic idea of a connection is of course that it is a preferred way of associating points lying in two ``nearby'' fibres.  It defines a splitting of the tangent space at any point in the bundle into the tangent space of the fibre (the vertical subspace) and a horizontal subspace---which is an ``infinitesimal bridge” to associated points in nearby fibres.   (Moreover, it does this in a way that respects the group action along the orbit: the `infinitesimal bridge' is  \emph{equivariant}.)

For our purposes, we only need to focus on two properties a connection can have: (a) being integrable, and (b) being unique, or unique subject to certain criteria. Both (a) and (b) will be relevant to our invoking connections, and related ideas like sections (and in Section 5: counterparts), in order to make comparisons between points in nearby fibres. 

Here we again see the two broadly different uses of `point’ that we noted in (2) at the end of Section \ref{intro}. For in the infinite-dimensional bundles of the preceding Subsection, `point’ means an entire field configuration; and so `comparing points’ means comparing global states of spacetime (or space). But Part I was about comparing {\em spacetime} points: i.e. elements inside the manifold $M$. Indeed, they were about ``identifying” (in less controversial language: putting in correspondence) with each other, two points in two isomorphic spacetimes---which are now configurations $\varphi_1, \varphi_2 \in \F$ that are in the same fibre. So it is clear that in order to relate connections in our infinite-dimensional bundles to the debates of Part I, we will need to ``descend” and ``look inside” the configurations $\varphi$, at how the field values are distributed over the points (i.e. the spacetime points!) in $M$. ``Looking inside” in this way will be the topic of Section \ref{sec:5cpart}. 

To discuss properties (a) and (b),   this Subsection will proceed in three stages. (1): We will first consider connections on the finite-dimensional bundles in familiar formulations of general relativity and gauge theories. (So in (1), it will be natural to read `point’ as 
 a point of spacetime or space, as in Part I.)  (2): Then we turn to connections on the infinite-dimensional bundles (field-spaces) $\F$ introduced in the preceding Subsection, though with technical details postponed to the Appendix (Sections \ref{sec:SdW_YM} and \ref{sec:SdW_GR}). (3): We complement (2)'s introduction to the general ideas with an intuitive example based on point-particle mechanics. This is called {\em best-matching}. It has the advantage of illustrating connections in the space of states, or configurations, like $\F$ for Yang-Mills and general relativity, while being finite-dimensional. (Again, there are more details in the Appendix (Section \ref{sec:SdW_particles}).) And finally after (3), we will end the Subsection with some general comments.

 But we first make some remarks about integrability that are true of both the finite- and the infinite-dimensional cases. Recall that a connection is called {\em integrable} if its distribution of horizontal subspaces is integrable. That is: through any point $p$ in the bundle there is a submanifold through $p$ (whose co-dimension is the dimension of the structure group) whose tangent space at any point is the connection’s horizontal space at that point. (This is the global definition: of course, one can also consider locally-existing submanifolds, and local sections.)

Of course, a connection may well not be integrable. For if it is, then for any point $p$ in the bundle, and any closed loop $\gamma$ in the base-manifold that starts at $p$’s projection in the base-manifold, the horizontal lift of $\gamma$ to the sub-manifold through $p$ will also be closed. That is, it will return to $p$: there will be no holonomy. Or in other words: the curvature of the connection is zero. And conversely, if the curvature is non-zero, i.e. there are non-zero holonomies, then the connection is not integrable.\\

(1): {\em The finite-dimensional case:---} We recall that for the finite-dimensional bundles
 in familiar formulations of general relativity and gauge theories, the connection represents a {\em potential}, and the curvature a {\em field-strength}. They each have a physical dimension, occur in the theory’s equations  and are different in different models of the theory.
In that sense, they are each `dynamical’.

More specifically, and philosophically: there is well-nigh universal consensus that in such theories, the field-strength counts as {\em physically real}. After all, it determines the motion of test matter located at the point in question. On the other hand, the gauge-freedom in the potential means there is debate about whether it counts as real---a debate that nowadays centres around the Aharonov-Bohm effect, especially for gauge theories.\footnote{We recall also that while a connection determines all these (possibly zero) holonomies around all closed loops---which are encoded at each point in the bundle by a curvature tensor (obtained, essentially, by differentiation of the connection)---the converse fails. That is: the curvature tensor does not determine the connection. (In abelian Yang-Mills theory, this follows already from the observation that two different connections that are gauge-related determine the same curvature. But in the non-abelian theory, two such connections may have different curvatures (since the curvature transforms in the adjoint representation). Nevertheless, in all cases: the curvature   need not determine the connection, since   non-gauge-related connections with the same curvature (even zero curvature)  may exist.}

But for our purposes, we do not need to enter this debate.\footnote{Since this entire Section treats general relativity and Yang-Mills theories on a par, we should mention that general relativity and non-abelian Yang-Mills theories have analogues of the Aharonov-Bohm effect (cf. e.g. \cite[Ch. 4]{Gomes_PhD} and references therein, and \cite{Overstreet_ABgrav} for a recent experiment).}
The reason is that our main points, here and  in (2) below, do not depend on whether we say that the gauge-freedom of the potential on these finite-dimensional principal fibre bundles renders that potential `physically unreal’.

The reason why our main points (here and in (2) below) do not thus depend arises from a general and uncontroversial aspect of how physicists, and we, think of the variety of different solutions of our theories. Namely: there being many different solutions represents {\em contingency}, or {\em happenstance}. One thinks of ``Nature choosing” a solution as a matter of contingent fact (often through a specification of initial and-or boundary conditions). And there is no onus on the theory (or on us) to somehow select or motivate that solution as ``preferred”.  

If one thinks of the variety of solutions in this way (as we do), then there is no question that in the familiar formulations of general relativity and gauge theories using a finite-dimensional principal fibre bundle (and of course, associated vector bundles):\\
\indent \indent \indent (i) the curvature (field-strength) is unique in the sense of being ``chosen by Nature”, i.e. being part of the solution that is actually physically  correct;   \\
\indent \indent \indent (ii) the connection (potential) is unique, up to a gauge choice, in the very same sense.

These points, (i) and (ii), will be enough for us. For as we will see in a moment, the corresponding points do {\em not} hold in the infinite-dimensional case. This will prompt the suggestion that in that case, which connection (and even which curvature) is used amounts to a theoretical choice, i.e. a human convention. This may at first seem ``unphysical” or ``subjective”. But rest assured: we will see in Section 5 that this conclusion fits very well with philosophers’ notion of counterparts.\footnote{\label{notanything}We stress that the choice, or convention, we intend here  goes beyond the familiar kind of gauge-fixing, such as choosing Coulomb gauge in electromagnetism, which the familiar fibre bundle formulations describe as choices of a section. We will see in (2) that the choice, or convention, we intend here is much more abstract;  and we will see in Section \ref{sec:5cpart} that there being a choice does {\em not} mean that ``anything goes”, i.e. that no reasons can be adduced for making one choice rather than another.} \\

\noindent (2): {\em The infinite-dimensional case:---}  We turn to connections on the infinite-dimensional bundles (field-spaces) $\F$ introduced in Section \ref{subsec:41bdle};   (again focussing on the properties of integrability, and uniqueness, that were labelled (a) and (b) in this Subsection's preamble). This is a large, technical and active subject. But in this paper, we will set aside all technicalities and restrict ourselves to introducing work by one of us (HG), together with others (with references): work that sets the scene for Section \ref{sec:5cpart}'s illustration of counterpart theory. We will leave technical details and examples to the Appendix (cf. Sections \ref{sec:SdW_YM} and \ref{sec:SdW_GR}).  Then  in (3) below, we will illustrate these ideas with a vivid finite-dimensional example, {\em best-matching}. It is a connection that  links the discussion
  to Part I's theme of comparing points of spacetime or space, rather than points of field-space. So adopting our metaphor: this connection ``descends" from field-space, and ``looks inside" the field-configurations at how the field-values are distributed over  points of spacetime or space.

One of us (HG) and coauthors (\cite{gomes_riem, GomesRiello2016, GomesRiello2018, GomesHopfRiello, GomesRiello_new}) show that on each of Section \ref{subsec:41bdle}'s bundles---of Lorentzian manifolds, and of field configurations of Yang-Mills theories (abelian and non-abelian)---one can define various connections; and state various  criteria that one can reasonably require the connection to satisfy.

 As in the finite-dimensional case, a connection-form is an appropriately smooth Lie-algebra-valued one form on $\F$, i.e. $\varpi: T\F\mapsto \fG$. Given $\xi\in \fG$ and $g(t)$ a curve  in $\G$ tangent to $\xi$ at $g(0)=\mathrm{Id}$, we define the \emph{fundamental vertical vector fields} 
 $\xi^\#_\varphi:=(\frac{d}{dt}\varphi^{g(t)})_{|t=0}$: these are the vector   fields that generate   at each $\varphi$ the tangent space to the orbit. Then, we can briefly state the conditions on the connection-form as:
\begin{eqnarray}
\varpi(\xi^\#)&=&\xi\label{eq:vert}\\
\bb{L}_{\xi^\#}\varpi&=&[\xi,\varpi]\label{eq:equiv},
\end{eqnarray}
where $\bb{L}_{\xi^\#}$ is the Lie derivative on field-space along ${\xi^\#}$, i.e. it is the infinitesimal action of  $R_g^*$: the pull-back of the form along the group action on $\F$; and $[\bullet, \bullet]$ is the commutator of $\fG$. (We here conform to the notation of \cite{GomesHopfRiello, GomesRiello_new}, using double-struck notation to distinguish differential geometric objects in the infinite-dimensional setting.) The finite version of \eqref{eq:equiv} is the perhaps more familiar $R_g^*\varpi=\Ad_g\varpi$, where $\Ad$ is the adjoint action of $\G$ on $\fG$.  

The second condition, Eq. \eqref{eq:equiv}, tells us the  connection must be equivariant, which  guarantees that the connection meshes with gauge transformations in the appropriate manner. The first condition, \eqref{eq:vert}, says that the action of the connection-form on the vertical vectors, that are generated by elements of the Lie algebra of the group, give back the corresponding Lie algebra generator. This condition is important for two reasons: (i) to relate the kernel of the connection form to a horizontal complement of the fibers (since the action on the vertical space already `fills up' the image of the map, and the kernel and the co-kernel sum up to the  domain); and (ii) because, jointly with the second condition, Eq. \eqref{eq:equiv}, it yields the grain of truth in the drag-along response (cf. Part I, Section 2.2.2), as we will discuss in Section \ref{subsec:522meas}.

In general, the connections---whether finite or infinite-dimensional; i.e. $\omega$ or $\varpi$---are of course not integrable, and so do not define sections. They are also not unique: more than one satisfies the criteria  that are reasonable to impose. But in the case of $\varpi$, they  are also not unique in the sense discussed at the end of (1) above: namely, unique in the sense of being dynamical, and so (putatively) ``chosen by Nature” in the solution, i.e. field configuration, that as a matter of happenstance is actually physically correct; as is $\omega$. 

In a bit more detail:  Gomes and coauthors mostly explore what they call the {\em Singer-DeWitt} connection. In terms of their general criteria, it is a very convenient choice. For firstly, it is induced by the kinetic term in the theory's action, as follows. In Yang-Mills theories, in a 3+1 split, the action can be written in terms of a kinetic term and a potential. The kinetic term is positive definite and defines a gauge-invariant inner product on field space; this inner product defines horizontal sections as those orthogonal to the group orbits. For more detail, cf. the Appendix,  Sections \ref{sec:SdW_YM} and \ref{sec:SdW_GR}.

Were we to employ a symplectic, i.e. Hamiltonian, formalism, we would replace field velocities by their momenta, but an equivalent reason can be given for the same connection. Namely, in the alternative symplectic framework, this same connection is also determined by symplectic orthogonality.   Thus in the Hamiltonian formulation of electromagnetism, the part of the electric field that is determined by the   instantaneous   distribution of charges is encoded in a Coulombic potential, and the Singer-DeWitt connection can be obtained as a projector onto the symplectically orthogonal complement of this potential; (see \cite{GomesButterfield_electro} for a philosophical introduction).  In the abelian Yang-Mills case, this connection is integrable; and it gives rise to the Coulomb gauge.  
In the non-abelian Yang-Mills case, it is not integrable; and does not give rise to a gauge-fixing. And in general relativity, it is again not integrable: though it is a natural choice, in particular by giving rise to the perturbative TT-gauge.

 Incidentally, it is unsurprising that these connections are in the non-abelian case not integrable: for they are everywhere defined, and had they been integrable, there would be a global section. But as is well-known, no such global sections can exist in the non-abelian theory: this is the Gribov obstruction \cite{Gribov:1977wm} that we mentioned in (2) of Section \ref{subsec:41bdle}.

So much  by way of introducing general ideas and results. We end this Subsection with a vivid, and tractably simple, example of a connection on a bundle whose points, i.e. elements of the bundle, are states of the physical system (rather than field-values at a spacetime or spatial point). Indeed this example is finite-dimensional, both in the fibres and in the base-manifold. \\

(3): {\em Best-matching: an illustration with Newtonian particles}:---  The system is a set of point-particles embedded in euclidean space; say $N$ of them, with masses $m_\alpha, \alpha = 1,2, ..., N$. So the ``absolutist" or ``Newtonian" configuration space of this system is $\RR^{3N}$, since for each particle, we need three real numbers to encode its position in absolute space. But suppose we are ``relationists" in the tradition of Leibniz and Mach: so that we regard two such absolutist configurations that differ by a spatial translation and-or a spatial rotation as physically the same---in physics jargon, as gauge-equivalent. Then it is natural for us to define the {\em relational configuration space} as the set of orbits (the quotient set) of $\RR^{3N}$ under the action of the euclidean group. Then as in the discussion in Section \ref{subsec:41bdle}, one can investigate the  ensuing bundle structure.
 Indeed, as in (2) of Section \ref{subsec:41bdle}: there  are reducible configurations, i.e. configurations that are fixed by an element of the euclidean group. For example, a collinear configuration (in which all the particles lie on a line) is fixed by any rotation about that line. So the orbits are of ``different sizes", and the most we can hope for is a stratified  orbit structure. In fact, this holds good (see \cite[Sec. III]{Littlejohn1997}).

And so we have a
 principal fibre bundle whose elements are absolutist configurations, whose structure group is the euclidean group, and whose orbits (or elements of the base-manifold) are relational configurations---forming a stratified manifold. It is a finite-dimensional ``cousin" of the stratified orbit, or quotient structure, of Riem($M$) (reported in  footnote \ref{ftnt:stab}).   Here, we say ``cousin'', despite a state of our system being a finite set of inter-particle distances $r_{\alpha\beta} (\alpha \neq \beta$) rather than a Riemannian metric across all of $M$, since in both cases ``there is no time"---the discussion is wholly about space, or matter in space. 

There is a very intuitive connection on this bundle. It is essentially due to \cite{Barbour_Bertotti} who call it {\em best-matching}. But Barbour and Bertotti did not think in terms of bundles, and so did not think of best-matching as amounting to a definition of a connection on this bundle. 
Here is  their original formulation of the best-matching procedure. First, (i) we think of each configuration, before one best-matches, as in its own copy of $\RR^3$; then (ii) one copy, together with its configuration, is  translated and-or rotated relative to the other copy, so as to minimise the mass-weighted average squared distance for each particle, labelled $\alpha$, as it occurs in one configuration and as it occurs in the other, i.e. so as to minimise $\Sigma_\alpha  m_\alpha  \dd r^2_\alpha$; then (iii) one says that each point in the first copy is to be identified with that point in the second copy, that it has been translated and-or rotated into coincidence with. 

Barbour and Bertotti suggested a vivid word for this identification or correspondence of points between copies of $\RR^3$. They say such a pair of points is {\em equi-local}. So to summarise: though best-matching is defined in terms of embedding the two configurations in a single copy of $\RR^3$, it also implies a notion of equi-locality between the two copies of $\RR^3$ in which the configurations are first given.   So using our jargon of `threading' (cf. Section 3.2 of Part I), one says that Barbour and Bertotti's equi-locality is a threading of points between two copies of $\RR^3$.

So the idea is: from a relationist perspective, an embedding of the two configurations into a single copy of $\RR^3$ that minimises $\Sigma_\alpha  m_\alpha \dd  r^2_\alpha$ is an embedding that {\em best-matches} the configurations. For it erases large contributions to the sum that would arise from embedding (the centre of mass of) one configuration at a great distance from the other: contributions that according to the relationist are an artefact of the mistaken postulate of an absolute space.

It can be shown that best-matching, so defined, is equivalent to  a connection: (cf. \cite{gomes_riem} for the case of spatial diffeomorphisms; and \cite[Ch. 5]{Flavio_tutorial} and \cite[Sec. 7.1]{GomesGryb_KK} for the particle case). The idea is that best-matching provides for a pair of absolutist configurations that are not congruent (and so are in different fibres/orbits of the euclidean group): a decomposition of the difference between them in to (i) a vertical gauge part (an element of the euclidean group which adjusts one of their embeddings in euclidean space, so as to minimise the functional $\Sigma_\alpha  m_\alpha  \dd r^2_\alpha$ of the two configurations) and (ii) a horizontal part, of which the minimum value of this functional gives a measure.   For more details, cf. the Appendix especially Section \ref{sec:SdW_particles}: it gives the explicit construction of the Singer-DeWitt connection, and its relation to best-matching as just described above.\footnote{By requiring orthogonality with respect to a fiber, best-matching ensures that horizontal (or best-matched) velocities have zero total angular and linear momentum. In \cite{GomesGryb_KK}, the link between best-matching and connections is made fully explicit. They also formulate   an extension of  these ideas  that is able to incorporate non-zero (but conserved) momenta.} 

For our purposes, the best-matching connection has another merit, in addition to being intuitive. Namely: the way that best-matching compares a pair of absolutist configurations ``looks inside" the spatial manifold (viz. $\RR^3$), in the sense discussed in the preamble to this Subsection. Indeed, best-matching does more than merely ``look inside", in the sense of turning our attention to the topic of Part I: i.e. the topic of ``identifying'' (or: putting in correspondence) spacetime or spatial points that occur in different models of our theory. For best-matching immediately  defines equi-locality as such an inter-model identification or correspondence of points. Namely: if two configurations are best-matched, i.e. embedded into a single copy of $\RR^3$ 
so that the embeddings minimise the functional $\Sigma_{\alpha}  m_{\alpha}  \dd r^2_{\alpha}$, then: the spatial points of that copy of $\RR^3$ are to be taken as the points of the configurations' spaces.

 To sum up the relevance of this illustrative example: these ideas of best-matching and equi-locality can be adapted to the infinite-dimensional, field-theoretic case. To make it explicit: we  embed two field configurations on the same smooth manifold, and minimise their difference according to a group of transformations and a notion of similarity. We thus replace:\\
\indent \indent (i) positions/configurations of a set of point-particles by the distributions of fields over spacetime or space; \\
\indent \indent (ii) $\RR^3$ by a differentiable manifold, $M$, in the case of metrics on spacetime or space;  and \\
\indent \indent (iii) the group of translations and rotations by the group of diffeomorphisms in the case of metrics over spacetime, or vertical automorphisms in the case of gauge theories; thus  the structure group (and also the base-manifold) is infinite-dimensional.\\

We  end  with three general comments about what we have done in this Section.\\
\indent (1): We have seen how a bundle of spacetimes---or in gauge theories, a bundle of field configurations---is a natural arena for making comparisons of spacetimes or configurations. But while a section, which provides a ``bridge’’ between the fibres that it intersects, is one way to make such comparisons, some bundles do not have global sections. So one naturally turns to the more general concept of a connection, which gives ``infinitesimal bridges’’, but might not give sections, i.e. might not be integrable.\\
\indent (2): In general, neither a section nor a connection in these bundles gives what Part I (especially Section 3.2) called a {\em threading} of the spacetime or spatial points in the copies of the manifold $M$ on which the elements (spacetimes or field configurations) $\varphi$ of the bundle are built. For although a section or a connection associates two ``nearby” elements $\varphi, \varphi'$, it might not do so by ``looking inside” the elements at the details of the distribution of field-values, and then defining the association in terms of a threading of points in the two copies $M$.\\
\indent (3): But at the end of the Section, we saw that best-matching is a connection that {\em does} look inside, and provides a threading of points in the different copies of its manifold $M$ viz.$\RR^3$. Thus best-matching leads us back to the topics of Part I; and so to counterpart theory . . .

\section{Non-isomorphic models: counterpart theory}\label{sec:5cpart} 

We now turn to how a section or connection on the bundle of spacetimes relates to {\em counterpart theory}.   We recall from Section 2.2.3 of Part I that this is the philosophical doctrine, due to David Lewis (\citep{Lewis1968}, \cite[p. 38-43]{Lewis1973}  and  \cite[Ch. 4]{Lewis1986}) that any two objects---in particular, spacetime points---in two different possibilities (in philosophical jargon: possible worlds) are never strictly identical. They are distinct: though of course  similar to each other in various, perhaps many, respects. And what makes true a proposition that the object $a$ could have had property $F$ (though in fact it lacks $F$) is---not that in another possible world, $a$ {\em itself} is $F$---but that in another possible world, an object appropriately similar  to $a$ is $F$. That object is called the {\em counterpart}, at this other possible world, of $a$.

There will be two main themes.\\
\indent  (1): First, a section or connection on this bundle {\em illustrates} both counterpart theory for spacetime points, and the threading of spacetime points (discussed in Part I's Sections 2.2.3 and 3.2 respectively).\\
\indent  (2): Second, a section or connection on this bundle  {\em generalises} counterpart theory for spacetime points. This will happen in two ways: the first is broadly philosophical, and the second mathematical. Namely:\\
\indent \indent (i):  A section or connection will define a counterpart relation, not just between spacetime points, but between entire spacetimes. There is a good precedent for this in the philosophy literature. Though we did not mention it in Part I,  Lewis and others propose not just that ordinary objects within two possible worlds can be counterparts, but also that vast extended regions of two spacetimes (endowed with their distributions of metric and matter fields), and even entire  spacetimes, can be counterparts of one another.\footnote{Lewis' own reason for this is that he accepts arbitrary {\em mereological fusions} of objects. As we said in Part I (cf. (3) in Section 2.1)): `mereology' is modern logic's name for the study of the part-whole relation; and a {\em fusion} of objects is, intuitively, the aggregate of them. More formally, it is the whole (i) that has them as parts, and (ii) whose every part overlaps at least one of the objects. So for Lewis, a possible world is itself an object, viz. the fusion of all the objects in it; cf. e.g. \cite[Section 4.3]{Lewis1986}.}\\
\indent \indent (ii): A section or connection will define a measure of how similar any two spacetimes (in general: in different fibres, and not on the section), i.e. any two elements  of the bundle, are. This measure is defined in terms of which elements of the bundle's structure group i.e. Diff($M$) project the spacetimes into the section. Besides, this measure has  a neat transformation property that reveals a grain of truth in what Part I called `the drag-along response' and associated claims (cf. Sections 2.2.2 and 3.1 of Part I).

We first give, in Section \ref{subsec:51prosp}, a prospectus for how we will discuss these themes.  Note that throughout this Section, we will for simplicity consider only vacuum spacetimes $(M,g)$ that are based on a single manifold $M$; (as we did in Section \ref{sec:4fibre}).

\subsection{Prospectus}\label{subsec:51prosp} 
First, we stress that all that follows applies equally to the bundle of field configurations in a Yang-Mills theory. So where we write `spacetimes', one could also read this as `field configurations'; and where we write `Diff($M$) is the structure group', one could also read this as `the group of gauge transformations is the structure group', and so on. But in view of this paper, and Part I, being about the hole argument, and for clarity of exposition, we will focus on spacetimes.

Second, we will for simplicity  discuss our themes mostly in terms of choosing a section of the bundle, rather than a connection. We agree that this is a mathematical simplification, since as discussed in Section \ref{subsec:41bdle}, the bundle in fact lacks global sections. We also agree that it is a philosophical simplification, since (as we will discuss below) a section can intersect two fibres representing spacetimes with very different physical contents (i.e. distributions of metric and matter fields: what Part I called `qualitative profiles'); and one might well feel that in some such cases, no point in one spacetime’s copy of the manifold $M$ is similar enough to a point in the other spacetime’s copy to be its counterpart. But we make these simplifications only to make the exposition clearer. All the proposals that follow could be stated in terms of a connection on, rather than a section of, the bundle;  so that only points (or extended regions) in two sufficiently similar spacetimes $\varphi = (M,g)$, i.e. two spacetimes in sufficiently close fibres, could be counterparts of each other.

The first theme of this Section ((1) in the preamble) is that a choice of a section in the bundle of spacetimes built on a fixed manifold $M$ can be  given {\em in part}  by a {\em pointwise specification}, within the spacetimes $\varphi$. That is: by an (infinitary) specification of the {\em threading} of points across the copies of $M$ in the different fibres. In terms of Section 4's metaphor: it is given by ``looking inside'' the spacetimes $\varphi$ and then threading the spacetime points in their copies of  $M$. Furthermore, we  think of a counterpart relation for points as specifying the threading.  We say `given {\em in part}', since the spacetimes being non-isomorphic means that the threading of points between spacetimes does {\em not} determine how the field values get altered, as well as re-distributed, between the spacetimes. Rather, it is the other way around: we would like the different distributions of fields to specify the threading between their copies of $M$.
 
This theme is clearest if we think of a curve $\gamma$, between two, non-isomorphic field configurations which  do not lie in the same section  (taken as the embedded sub-manifold). That is, we are given a smooth map of a real interval, say [0,1], whose image or range is a 1-parameter family of spacetimes $\varphi$, such that $\gamma(0)=\varphi_1$, and  $\gamma(1)=\varphi_2$. Since the initial and end points of the family are not isomorphic, this family is {\em not} defined by a 1-parameter family of diffeomorphisms of $M$. Along $\varphi_2$'s orbit, $\mathcal{O}_{\varphi_2}$, there will be a single distribution, $\varphi_2'$, such that its copy of $M$ is threaded by the identity diffeomorphism to $\varphi_1$'s copy of $M$. That distribution, $\varphi_2'$, lies on the same section as $\varphi_1$. All other elements of $\mathcal{O}_{\varphi_2}$ will be threaded by the drag-along from $(M, \varphi_2')$. We could project $\gamma$ onto the section, in which case a 1-parameter family of diffeomorphisms would partly specify the curve along the section.  To sum up: we can envisage a curve lying in the section as involving (being partly specified by) a threading of spacetime points by the diffeomorphisms---but this threading is {\em not} by drag-along. Thus we return to Part I's third and fourth examples of threading, viz. Geroch’s definition of limits of spacetimes and quantum reference frames (Sections 3.2.3 and 3.2.4).

 On the other hand, the second theme of this Section ((2) in the preamble) is that, relative to a section (or connection), we can think of entire spacetimes as counterparts. As we mentioned: this accords with some of the philosophical literature, which proposes that entire spacetimes can be counterparts. Our main comment about this theme will be that the section defines a `measure' of how similar any two (models of) spacetimes are. This `measure' is defined in terms of which elements of the bundle's structure group, i.e. Diff($M$), project the spacetimes into the given section. Besides, this `measure' has a neat   transformation  property.

We stress that in the case of sections, the `measure' of similarity is not a measure in the sense of measure theory; nor need it be some sort of minimum distance, or a functional of such a distance. But here we will see an advantage of using infinitesimal sections, i.e. connections, so as to define counterparts in neighbouring fibres. Such counterparts will minimise a well-defined measure of difference between spacetimes, and so our measure of similarity will indeed encode an intuitive  sense of similarity.\footnote{\label{sensible} {Though this measure of counterparthood can be given an account in terms of similarity in the case of infinitesimal connections, here, for purposes of illustration,  we will not dwell on the difference. Note that although this neat measure of counterparthood is for entire spacetimes (theme (2)), we envisage that for ``philosophically sensible'' sections or connections (cf. Section \ref{subsec:521dissim}),  counterparthood for spacetimes is determined by all the counterparthood relations between their points (theme (1)).}}
 
It will be clearest to develop these themes by focussing on the more novel of them, i.e. (2). For it will be clear how the treatment of (1) gets subsumed by (2). So we will first, in Section \ref{subsec:52sptcpart}, state the main idea of (2). Then in Section \ref{subsec:521dissim}, we will admit that for an arbitrary section, the spacetimes that it rules to be counterparts can represent very different physical states. That is, they can be on two fibres representing possibilities that are qualitatively (`no matter which point is which') very different.   Finally, Section \ref{subsec:522meas} will be more positive: we define (even for an arbitrary section) a measure of similarity of spacetimes, with a neat   transformation  property.

\subsection{Spacetimes as counterparts}\label{subsec:52sptcpart}

We briefly recall the notation from Section \ref{subsec:41bdle}. 
 We denote the bundle of spacetimes built on the manifold $M$ by $\Phi$; and we call an element of it, i.e. a Lorentzian manifold, $\varphi$. We write each gauge-equivalence class, i.e. fibre or orbit,  as $[\varphi]$, and to stress its being an orbit, as $O_{\varphi}$.  
A section, $\sigma$ is a choice of a state $\varphi'$ in each orbit $O_{\varphi}$: so $\varphi'$ is the representative, according to this section, of $O_{\varphi}$. (Again, this is as in gauge theory: where a section is a gauge-fixing.) 

We will not need to distinguish in notation between two different notions of a section:\\
\indent \indent (i) the function from the bundle’s base-manifold (i.e. the set of orbits/gauge-equivalence classes) into the bundle, which is usually written $\sigma$; and \\
\indent \indent (ii) the range of this function, which is the embedded sub-manifold of the bundle. This is often written $\cal F$: think of `$\cal F$’ as standing for `folium’, i.e. leaf. ($\cal F$ is also used for a real-valued function on the bundle for which the folium is the level-surface with value 0. That is: iff $\varphi$ is in the section, ${\cal F}(\varphi) = 0$.)\\ 
So we shall always say `section’ and write $\sigma$. 

Our key proposal is to extend the familiar idea that a section is a choice of representative states, one for each physical state. Namely: to the proposal that a section is a choice, for any state, and for any fibre/orbit, of which state in that fibre corresponds to---we shall say: is the counterpart of---the given state. That is, we propose: 
\begin{quote}
A section stipulates, for each state $\varphi$ (not necessarily in the section), and each orbit $O_{\varphi’}$  (in general distinct from the orbit of $\varphi$): which state,  $\varphi’'$ say, in $O_{\varphi’}$ corresponds to (we say: {\em is the counterpart of}) $\varphi$. The stipulation is the obvious one: the counterpart state $\varphi’'$ is stipulated to be the element of the section within $O_{\varphi’}$.
\end{quote} 
So the intuitive idea is that in order to go from the arbitrary given state $\varphi$ to the state in $O_{\varphi’}$ that is its counterpart, one projects $\varphi$ vertically along its fibre into the section, and then travels horizontally across the section to $O_{\varphi’}$, and thus arrives at the counterpart state $\varphi’'$. 

Note that in these last two paragraphs, the word `state' could  be replaced throughout by `spacetime'. Similarly, for the discussion to follow. But we will continue, mostly, to say `state', not least because this signals that our proposal applies---all our considerations apply---equally to Yang-Mills theories.   

We will now develop this proposal. 
We do so in two Subsections. First, Section \ref{subsec:521dissim} will register a note of caution. For we will admit that since the general idea of a section is formal, what a section stipulates as the counterpart state  to the given one may be very dissimilar from it.\footnote{Besides, as discussed in Section \ref{subsec:51prosp}:  even if the corresponding state is similar, the section does not itself specify a threading of spacetime points.} But we will maintain that this allowance of dissimilarity is no disadvantage. 

Then Section \ref{subsec:522meas} will be more positive. We will show how---by considering not only projecting $\varphi$ vertically along its fibre into the section, but also vertical projections along the ``destination fibre” $O_{\varphi’}$---we get a natural definition (relative, of course, to the choice of section) of a {\em measure} of ``how much" one state is a counterpart of another.

\subsection{The counterpart state can be dissimilar}\label{subsec:521dissim}

It is clear that, since our idea invokes {\em any} section of the bundle, the state in a fibre $O_{\varphi’}$ that is the counterpart of a given state $\varphi$ can be very dissimilar from $\varphi$. 

In fact, there are two points here. We will state them, and then urge that allowing such dissimilarities is no disadvantage. That is: this allowance  is no objection to our using sections of the fibre bundle of spacetimes to make comparisons of non-isomorphic spacetimes.    

First, the dissimilarity can be regardless of how we   might choose to thread (in the usual jargon: trans-world identify) spacetime points: regardless of which point is which, in the comparison of the manifolds. For example, the state $\varphi$ might have an  intuitively uniform geometry---it might be a Lorentzian manifold $(M,g)$ with $g$   intuitively close to the Minkowski metric---while all the states in another fibre $O_{\varphi’}$, i.e. all the mutually isometric Lorentzian manifolds in that different fibre, are very ``wrinkly’’. So no matter where the section intersects the fibre $O_{\varphi’}$, the manifold, $(M,g’)$ say, which {\em is} that intersection will have a very different geometry from Minkowski. This means that there is no judicious way to thread (trans-world identify) points between the two copies of $M$---one in the given $\varphi \equiv (M,g)$ and the other in  $(M,g’)$ i.e. the section’s intersection with $O_{\varphi’}$---that will make the qualitative geometric profiles (i.e. distributions of geometric attributes) within the two copies be similar (i.e. have equal, or nearly equal, field-values).\footnote{ We say `$g$ intuitively close to the Minkowski metric', instead of `$g$ the Minkowski metric', so as to set aside the subtleties about extending a slice between different strata, which we discussed in (2) of Section \ref{subsec:41bdle}.}

 Second, even if the initially given spacetime $(M,g)$ and its counterpart relative to the section $(M,g')$ {\em do} have similar qualitative geometric profiles: nevertheless, some assumed or mathematically natural threading of points between $(M,g)$ and $(M,g')$\footnote{\label{care}  Although a threading is of course {\em not} defined by a section, it might be defined by a projection of $(M,g)$ in to the section, followed by a curve in the section, i.e. a 1 parameter family of spacetimes, ending in $(M,g')$, if we suppose that the curve's definition includes a 1 parameter family of diffeomorphisms (as mentioned in Section \ref{subsec:51prosp})---though of course the spacetimes are {\em not} given as drag-alongs by the diffeomorphisms. } might well not ``follow the spirit" of genuine similarity, i.e. of pairing together equal, or nearly equal, field-values. The obvious case of such a threading---assumed or mathematically natural, but violating the spirit of similarity---is the identity map on the base-set of the manifold $M$. In such a case (i) there {\em is} a judicious way to thread points---other than by identity map---between the two copies of $M$---one in the given spacetime $\varphi \equiv (M,g)$ and the other in {\em some member} of the fibre $O_{\varphi’}$---that makes the points thus threaded to each other have similar qualitative geometric profiles;  but (ii) the section does not intersect the fibre $O_{\varphi’}$ in that member. 

To sum up these two points: a section is a choice, for each physical state  (global distribution of geometric attributes) $O_{\varphi}$, of a representative spacetime $\varphi' \in O_{\varphi}$. 
But there need be no---and even a curve in the section need not suggest a---threading of spacetime points between the representative spacetimes that keeps track of equal, or nearly equal, field-values. Here is a toy-model spatial example to illustrate these two points. For the second point, it will take the identity map on the manifold $M$ as the assumed threading.\\

\noindent {\em A Spatial Example}:---  Take the underlying spatial manifold $M$  to be $S^1$, which we take as $U(1) = \{ \exp(i \theta): \theta \in [0, 2\pi ) \}$. The only field is a real scalar.  We take field configurations that differ by the action of a rotation in $U(1)$ to be physically the same: so the idea is that the identity of points is gauge. So the structure group is $U(1)$. We will consider a state $\varphi$, in which this field is mostly constant, say 0, but has a sharp peak around $\exp(i \pi) = -1$, i.e. a peak with a small support in $\theta$, centred around $-1 \in S^1$. 
So the other states in the orbit of $\varphi$ have a congruent sharp peak centred around various $\exp(i \theta)$, with $\theta \neq \pi$.

To illustrate the first point above---that a section can stipulate a qualitatively dissimilar state as the counterpart of our given state---we need only consider a physical state (fibre, orbit) with, say, seven peaks rather than one, distributed around the circle $S^1$; and a minimum value in some valley between the peaks that is much greater than 0. Clearly: no matter where exactly these peaks are located around $S^1$---no matter where the section intersects this fibre---the two physical states are very dissimilar. No judicious threading (in the usual jargon: trans-world identification) of points between the two copies of $S^1$ can make them similar.

To illustrate the second point above, we choose a physical state (fibre, orbit) that is qualitatively very similar to our given state $\varphi$, e.g. with a single peak nearly congruent to the peak in $\varphi$. But then we consider a section that intersects this fibre ``perversely”, i.e. intersects the fibre in a state that does not ``place" the peak over the same points as in our given state $\varphi$.   (The phrase `the same points' signals our invoking the identity map on the manifold $U(1)$  as the assumed threading. This will be in play in the rest of this Example.) 

In a bit more detail: consider a specific physical state (fibre, orbit) $O_{\varphi’}$ that is similar---with the field-value on most of $ S^1$ being the same constant, i.e. 0, but with a less sharp (so: not congruent) peak somewhere on the circle. Suppose our chosen section $\sigma$ ``happens" to intersect $O_{\varphi’}$ at a state, $\varphi’'$, where the less sharp peak is centred round $\exp(i \frac{\pi}{2})$.

That is: $\sigma$ happens to stipulate that the representative (orbit-element) that is the counterpart of $\varphi$---according to $\sigma$---is a state that (i) rotates $\varphi$’s sharp peak anti-clockwise by $\frac{\pi}{2}$, and (ii) widens it a little to be less sharp (so as to coincide with the peak of $\varphi’'$).

We thus see that the section $\sigma$  violates, by a quarter-turn,  the spirit of genuine similarity. For  comparing these non-isomorphic models in terms of similarity would suggest ``trans-world identifying” points below the two peaks.  One might say: `although the peaks are not congruent, we should still think of the two peaks as occurring in the same approximate region, e.g. with the summit at the same place in the circle. In other words, we should identify ``same place” in terms of the values of the scalar field, allowing for approximate matching.' But our section $\sigma$ does {\em not} do so. That is of course unsurprising, since we chose it arbitrarily.

It is also worth noting the changes between states given by following curves lying in the section. A curve $\gamma$ lying in the section $\sigma$ is a 1-parameter family of states. (More precisely: $\gamma$'s image is such a family.) We take it as  given by a 1-parameter family of diffeomorphisms of $S^1$, together with 1-parameter family of deformations of the scalar field. (We say `deformations' because the field’s ``profile’’ or ``shape’’ varies along the curve, since the points are physically different---they all lie in different orbits.)

Thus in our Example: there are curves in the section that connect $\varphi$ and $\varphi’'$. Some curves make the required changes in states  very straightforwardly. For example: they rotate the sharp peak anti-clockwise by $\pi/2$, while also smoothly broadening the sharp peak so that at the end of the curve it is exactly the peak of $\varphi’'$. And other curves make the required changes in a more complicated way. For example: the sharp peak gets rotated anti-clockwise by $5\pi/2$ (i.e. a full loop around $S^1$), while being smoothly deformed so as to eventually give exactly the peak of $\varphi’'$. And this smooth deformation could be ``inefficient”: the curve in the section might first smoothly make the peak yet sharper, and only then, after the sharpening, broaden it so as to eventually coincide with $\varphi’'$.\\ 

So much by way of an example: we now sum up this discussion. An arbitrary section will in general provide an unintuitive standard of which pairs of states in the various pairs of orbits are counterparts of each other.  Here, `unintuitive'  means the pairs of states are in general not similar in their profiles of field-values, nor in their assignments of field-values to points.
But we maintain that this allowance of dissimilarity is no objection to our using sections of the fibre bundle of spacetimes for making comparisons of non-isomorphic spacetimes. It is simply an immediate consequence of our having so far put no constraints on the choice of section---and we are at liberty to add such constraints, and so make the counterpart relations (the correspondences) given by sections track similarity better.\footnote{\label{ftnt:threecomms}Three further comments, in increasing order of mathematics. (i): For example, such constraints might make counterparthood for points and for regions mesh appropriately; cf. footnote \ref{sensible}. (ii): Such constraints might also make use of `observables'
  by which, generically, spacetime points can be uniquely labelled (e.g. Komar observables  \cite{Komar_inv}, mentioned in Part I, at footnote 20). (iii): Though we have spoken for simplicity of sections, really one should make sense of these constraints on similarity in terms of infinitesimal sections, i.e. connections: (cf. again footnote \ref{sensible}).  This is one advantage of using connections in place of sections: connections yield a measure of similarity between spacetimes that can encode information about natural notions of distance or difference between the spacetimes. 

 That is, assuming any equivariant complement of the tangent spaces to the orbits can be given by an orthogonality condition for some suitable inner product on $\F$, we can interpret a pair of counterparts in neighbouring orbits as minimising a corresponding notion of difference between spacetimes. Moreover, we can constrain the choice of inner products by suitable requirements on how we would like to compare the spacetimes, e.g. locally, ultralocally (that is, without taking derivatives of the metric), etc.  In more detail, one such constraint is that the connection be given by orthogonality with respect to a supermetric, $\bb G$ (see  the Appendix), but demand that $\bb G$ be \emph{ultra}local: i.e.  expressed as an integral on the local components of the field space vectors, without derivatives. For example, for the space of spatial metrics Riem (cf. Section \ref{sec:SdW_GR}), it is easy  to see (as first shown by DeWitt) that this ultralocality condition is only satisfied by a one-parameter group:  
\be \bb G( \bb X, \bb Y)_g:=\int \sqrt{g}\, (g^{ac}g^{bd}-\lambda g^{ab}g^{cd})(x) \bb X_{ab}(x) \bb Y_{cd}(x), 
\ee
which,  in dimension 3, are supermetrics (i.e. positive-definite) only for $\lambda<1/3$. (See also footnote \ref{ftnt:dewitt}.)  In the Yang-Mills case, there is similarly a one-parameter family in the non-Abelian case (arising from two inner-products on the Lie algebra), and a single such metric in the Abelian case. Cf. the discussion in (2) and (3) of Section \ref{subsec:42connn}; and for details, the Appendix.}

Essentially this point is a familiar one in gauge theory. In that context, it is often pointed out that the choice of a gauge (i.e. a section of field-space) is in principle arbitrary, in the sense that the physics can be accurately stated in any gauge. But of course that arbitrariness or flexibility is entirely compatible with there being good reasons (e.g. of calculational convenience, or of theoretical clarity) to choose one gauge rather than another (see footnote \ref{notanything}). And these reasons need not be specific to a single problem or group of problems. For example: in the Hamiltonian treatment of electromagnetism, there are theoretical reasons to favour the Coulomb gauge that are wholly general, and so apply to any problem (\cite{GomesButterfield_electro}).

\subsection{A measure of similarity---relative to a section}\label{subsec:522meas}

We now set aside the concerns Section \ref{subsec:521dissim}.  In this Section, we describe how relative to a choice of section $\sigma$ (so: no matter how unintuitive its verdicts of counterparthood might be) there is a natural `measure' of how similar, according to the section, two states (spacetimes) $\varphi, \varphi’ \in \Phi$ are. These can be {\em any} two spacetimes (on the fixed manifold $M$): neither need be on the section. 

 As we pointed out at  the end of Section \ref{subsec:51prosp} (and in accord with the closing comments of Section \ref{subsec:521dissim}): for a section, this `measure' will not be a measure in the sense of measure theory; nor need it be some sort of minimum distance, or a functional of such a distance. But in it infinitesimal version, i.e. for a connection, it can be such a functional:  see footnote \ref{ftnt:threecomms}). Nonetheless, this `measure' is natural in that:\\
\indent \indent (i) it is determined by the section (for any section); \\
\indent \indent (ii) it encodes the similarity as, roughly speaking, the element of the diffeomorphism group that, combined with some ``horizontal travel” across the section, would transform the spacetime $\varphi$ into the spacetime $\varphi’$;\\
\indent \indent (iii) for two states (spacetimes) that both lie in the section, the measure is  what we might call `utter similarity'  i.e. maximal---represented mathematically by the identity element of the group, $Id \in$ Diff($M$): \\
\indent \indent (iv) the measure has a neat property: it   transforms  in an appropriate way under the action of a single group element (i.e. a change of section made by translation with a single group element); \\
\indent \indent (v)  finally, the measure and its properties apply to any principal fibre bundle.       

Namely: we measure the similarity of two states in terms of how much they need to slide vertically along their respective orbits, in order that their images  under the sliding both lie in the section. It is convenient to define this measure ---now dropping the scare-quotes---as the product of one structure-group element, with the inverse of the other.

To say this exactly, we need some notation. Since as just noted in (v) above, the ideas work for any principal fibre bundle: we write the structure group as $\G$ with elements $g$; (so $g$ is a group element, not a Lorentzian metric!). 
 We write the (right) action of the structure group $\G$ on the entire bundle (the field space) $\Phi$ with elements  $\varphi$, in the usual way, as a superscript. The action is:
\begin{equation}\label{Gaction}
 (\varphi \times g) \mapsto \varphi^g \in O_{\varphi}.
\end{equation}

A section $\sigma$ defines what we might call a ``ticket map”. This map assigns to each $\varphi \in \Phi$, the element in $\G$ that transforms $\varphi$ to its projection in the section, i.e. to $O_{\varphi}$’s intersection with $\sigma$.  We write this element, the value of the ticket map, as $g_{\sigma}(\varphi)$. We say ``ticket’’ because the element $g$ is the ticket that projects $\varphi$ vertically along its orbit to the section.  So we write the ticket map as: 
\begin{equation}\label{intoorbit}
\varphi \mapsto g_{\sigma}(\varphi) \in \G; \;\;  \mbox{with} \; \;  \varphi^{g_{\sigma}(\varphi)} \in O_{\varphi} \cap \sigma.
\end{equation} 

There is a straightforward relation between the values of the ticket map for $\varphi$ and for its transform $\varphi^g$, i.e. between $g_{\sigma}(\varphi)$ and  $g_{\sigma}(\varphi^g)$. We can write it in two equivalent ways. Namely:\footnote{As stated in Section \ref{subsec:51prosp}, we have here made the simplifying assumption that the section cuts every orbit only once; that is, in the nomenclature of \cite[Sec. 3]{Samediff_1b}, that the corresponding conditions determining the section satisfy `universality' and `uniqueness'.} 
\begin{equation}\label{equivarianceofticket}
g_{\sigma}(\varphi^g) = g^{-1}g_{\sigma}(\varphi) \;\; ; \;\; \varphi^{g_{\sigma}(\varphi)} = (\varphi^{g})^{g^{-1} g_{\sigma}(\varphi^g)} \, .
\end{equation} 
This relation, eq. \eqref{equivarianceofticket}, will imply that the measure of counterparthood we are about to define has a neat transformation  property (cf. eq. \eqref{covarycpart}).\footnote{As an example, take Coulomb gauge in Hamiltonian electromagnetism. The section $\mathcal{F}$ is defined by the regular values $\nabla^i A_i=0$. The `ticket map' is given by $g_\sigma(A)=-\nabla^{-2}(\nabla^i A_i)$: which, as is easy to verify, satisfies, for $A^g_i=A_i+\nabla_i g$, the equation: $g_\sigma(A^g)=g_\sigma(A)-g$. When we `dress' $A_i$ with $g_\sigma(A)$, as in the second equation of \eqref{equivarianceofticket}, we obtain the projection to $\mathcal{F}$, given by $A^{g_\sigma(A)}_i=A_i-\nabla_i(\nabla^{-2}(\nabla^j A_j))$: which by the transformation property of the ticket map satisfies the second equation of \eqref{equivarianceofticket}. In other words, the dressed or projected gauge potential is gauge-invariant. }

We now define the measure of how much two states, $\varphi_1$ and $\varphi_2$, need to slide vertically along their respective orbits, in order to both lie in $\sigma$, as the product of two ``tickets”. But for the second ticket in this product, we take the inverse. This expresses the idea of travelling from the first state $\varphi_1$ ``down” into the section $\sigma$, then travelling horizontally across the section to the orbit of $\varphi_2$, and then vertically ``up” out of the section to arrive at $\varphi_2$.

That is, we define: For any section $\sigma$,
\begin{equation}\label{defcpart} 
	\boxed{\quad\phantom{\Big|}
{\rm Counter}_{\sigma}(\varphi_1, \varphi_2) :=  g_{\sigma}(\varphi_1)g_{\sigma}(\varphi_2)^{-1} \; \quad}.
\end{equation}
And note also that 
\be\label{eq:counter_comp} {\rm Counter}_{\sigma}(\varphi_1, \varphi_2){\rm Counter}_{\sigma}(\varphi_2, \varphi_3)={\rm Counter}_{\sigma}(\varphi_1, \varphi_3)
\ee

So our proposed measure, Counter$_{\sigma}$, of how much any two states are counterparts, relative to the section $\sigma$, is itself an element of the structure group: in our general notation $\G$, but in the case of interest to us, Diff($M$). Of course, `measure’ here has nothing to do with measure theory. Nevertheless, the way that Counter$_{\sigma}$ encodes degrees of similarity, as judged by $\sigma$, by an element of $\G$ is natural, in that:\\
\indent \indent  (a) if $\varphi_1, \varphi_2 \in \sigma$ (i.e. the states are already in the section), then Counter$_{\sigma}$($\varphi_1, \varphi_2$) is the identity element of $\G$; \\
\indent \indent  (b) $\G$ is a topological (in particular, Lie) group, so that it makes sense to talk of being near or far from the identity element.

Thus we see how the definition of Counter$_{\sigma}$, eq. \ref {defcpart}, immediately implies the features listed as (i) to (iii) above. We now turn to feature (iv).

For this, we simply apply the first equation of eq. \ref{equivarianceofticket} (which is equivalent to the second) to the expression Counter$_{\sigma}(\varphi_1^g, \varphi_2^g)$, getting,
 for any section $\sigma$ and any $g$: 
\begin{equation}\label{covarycpart} 
{\rm Counter}_{\sigma}(\varphi^g_1, \varphi^g_2) = g^{-1}g_\sigma(\varphi_1)(g^{-1}g_\sigma(\varphi_2)^{-1})= g^{-1}\,{\rm Counter}_{\sigma}(\varphi_1, \varphi_2)g\; .
\end{equation}
That is: our counterpart measure, eq. \ref{defcpart},  transforms by conjugacy
under the action of each element $g \in G$.

 We end this Section with two further features---we claim: advantages---of our measure of similarity, i.e. of \eqref{defcpart}, and its two consequences \eqref{eq:counter_comp} and \eqref{covarycpart}. Broadly speaking, the first feature, (1), is mathematical; the second, (2), is philosophical. \\

(1): Equation \eqref{defcpart} has a consequence that is worth noting. Two models $\varphi_1, \varphi_2$  that lie in the same orbit will always be related by the  unique isomorphism that connects them.\footnote{\label{stab2}{Assuming that the orbit does not have stabilisers, the ticket map is unique, and so is the counterpart relation. If that orbit has stabilisers, the ticket map is not unique, and so there will be an ambiguity: the counterpart relation is unique up to stabilisers.}} That is: if $\varphi_2 =  \varphi^g_1$, then ${\rm Counter}_{\sigma}(\varphi_1, \varphi_2) = g$, even if neither $\varphi_1$ nor $\varphi_2$ lie in the section $\sigma$. 
For with the definitions above,
\begin{equation}\label{same}
{\rm Counter}_{\sigma}(\varphi_1, \varphi^g_1) = g_{\sigma}(\varphi_1)(g^{-1}g_{\sigma}(\varphi_1))^{-1}=g.
\end{equation}

 Furthermore, \eqref{same}, and the preceding equations, have consequences for  the traditional   (and our Part I's)  topic of the counterparthood of single points, rather than entire spacetimes. So let us now focus on the case of Lorentzian metrics and diffeomorphisms: so that the $\varphi$ are now spacetime metrics $g_{ab}$ (but we will omit indices in what follows) and the group elements $g\in \G$ are now diffeomorphisms $d \in$ Diff($M$).

Given two models, $(M,g_1)$ and $(M,g_2)$, and some choice of $\sigma$, we obtain the diffeomorphism $\tilde d:={\rm Counter}_{\sigma}(g_1, g_2)$. Thus given point $p$ in model $(M,g_1)$---we will use a shorthand for such a doublet: $(p, g_1)$---its unique counterpart (in the sense, of course:  `counterpart of points, not of entire models') at $(M, g_2)$, is $\tilde d(p)=:q$. Or in the shorthand notation, the counterpart is $(q, g_2)$. 

Now suppose that $g_2$ is isometric to a third metric, $g_3$, i.e. that $d^*g_2=g_3$ for a unique $d \in$ Diff($M$). Then by \eqref{same}, given any point $r$ in model $(M, g_2)$, its counterpart in $(M, g_3)$ is $d(r)$.  Then it follows from \eqref{eq:counter_comp}
that:
\be  {\rm Counter}_{\sigma}(g_1,  g_3)=d\tilde d,
\ee
and so  the counterpart of  $(p, g_1)$   in  $(M,g_3)=(M,d^*g_2)=(M,g_2^d)$ is $d(q)$, or, in shorthand notation, it is $(d(\tilde d(p)), g_2^d)=(d(q), g_2^d)$.\footnote{ Here of course, the superscript notation $g_2^d$ just denotes the action of Diff($M$) on $F$; cf. \eqref{Gaction}.} In other words,  $(p, g_1)$ is the counterpart of  $(q, g_2)$ iff it  is the counterpart of $(d(q), g_2^d)$.
Besides, this  property is independent of which section  we choose.\\

(2): Finally, we turn to the promise we made in (2) of Section \ref{subsec:42connn}  and in Part I, that this Section's framework of counterparts for spacetimes  would reveal a grain of truth in what Part I called `the drag-along response', and in the claim we there rebutted, that the drag-along response is mathematically compulsory; (cf. Part I, Sections 2.2.2 and 3.1 respectively). 

What  we have in mind is not just the mathematical fact that there is a unique\footnote{Again, we note that uniqueness holds only generically: recall that models with automorphisms (reducible configurations) spoil uniqueness. See footnotes \ref{ftnt:stab} and \ref{stab2}. More specifically:  entire models will, under our assumptions, always have unique counterparts, even if they are reducible. But if one of the counterpart models is reducible, the spacetime points have duplicates. So  one spacetime point in a model of one isomorphism class will have more than one counterpart in the (unique counterpart) model with stabilisers. \label{ftnt:stab_counter}} isomorphism between two elements of a fibre. After all, this mathematical fact obviously stands ``on its own feet", irrespective of any mention of counterparts. Our point is, rather, that (i) thinking of the bundle interpretatively, in terms of counterparts---with the connotations of similarity and flexibility that that word now has in philosophy---and (ii) defining (degrees of) counterparthood as elements of Diff($M$), provides an interpretative context in which this unique isomorphism earns the name of being a ``trans-world identification". That is: by putting the unique isomorphism in a single framework together with other diffeomorphisms that 
are not isomorphisms, this context supports the idea that two spacetime points that the isomorphism relates have the best possible claim to ``correspond".  That is: the best possible claim to be the ``same" as each other---to use, for once, that  suspect  word `same'! Thus we come full circle, back to the basic themes that launched the hole argument, and our endeavours here and in Part 1.

Thus, although our counterparts rescue this grain of truth from the drag-along response, they are not subject to our criticisms of that response. First, by weakening the `identification' endorsed by the drag-along to `best possible correspondence', we are not subject to the problem posed by symmetric  models (see (3) in Section 2.2.2 of Part I, and footnote \ref{ftnt:stab_counter}). Second, we can detach diffeomorphisms from their role of inducing the `best possible correspondence': diffeomorphisms are allowed to map between points with different qualitative profiles, so that they don't preserve physical facts pointwise. Thus diffeomorphism symmetry is `global': it is the totality of qualitative profiles of all the points of a model that is the same as that of diffeomorphism-shifted model---and non-trivial: it can have non-trivial
consequences, e.g. through Noether's second theorem (Section 3.2.2 of Part I).

Lest the reader feel there is some sleight of hand, let us be as clear as possible about the interpretive assumptions that salvage this grain of truth from Part I’s criticisms of the drag-along. The point is this: independent of our endorsement of what Part I called `Sophistication’, using general relativity will often, if not always, require us to choose a single representative model from each isomorphism class. If such choice is made consistently and the choice is smooth in the space of models, then one gets a section of the fiber bundle. What we have shown above (cf. especially  (1)) is that any such choice not only allows a definition of counterparts between spacetime points in non-isomorphic spacetimes. but also: those counterparts also give an equivalence relation between spacetime points (in isomorphic and non-isomorphic models) that respects the drag-along.

\section{Conclusion}\label{concl}

Let us briefly summarise what we have done in this two-part paper. In Part I we reviewed, from a philosophical perspective, the hole argument, which is about the ``identification’’ of points between isomorphic spacetimes. Here in Part II, we proposed a fibre bundle of spacetimes, with the diffeomorphism group for a fixed spacetime manifold $M$ as structure group, as a framework for comparing {\em non}-isomorphic spacetimes: using sections, and locally connections, as ``bridges”  between elements in different fibres of the bundle. We stressed that this proposed framework is strongly analogous to existing field-space formulations of gauge theories. In particular, the technicalities about our proposal’s connections have been developed primarily for gauge theories;  as  expounded  in the Appendix.

For this paper’s endeavour, the main contributions of Part I were two ideas. Namely: the philosophical idea of points being counterparts; and the more technical idea of threading, i.e. the idea that mathematics and physics sometimes associate, or ``identify”, points in different manifolds by a mapping {\em other than} isomorphism. These ideas were combined with the idea of a fibre bundle of spacetimes, in this paper's Section \ref{sec:5cpart}. There, we saw how to define a measure of similarity or counterparthood (relative to a section) for entire spacetimes, in terms of the bundle’s structure group; and we even showed that there was a grain of truth in doctrines that our Part I had rebutted, to the effect that isomorphism is the only good way to associate, or ``identify”, points. 
 Thus we hope that in the philosophical controversies about the hole argument, some peace might break out . . .

Of course, various questions about our proposed framework remain to be explored. Two obvious ones, from a philosophical and a physical perspective respectively, are as follows.\\ 

\indent \indent (a): Which of the various versions of counterpart theory, that are to be found in the logico-philosophical literature (whose details we have here ignored), fits best with our proposed comparisons of spacetimes, and of spacetime points, using sections and connections on a fibre bundle?

\indent \indent (b): Which of the various threading schemes for non-isomorphic spacetimes that are to be found in the mathematical physics literature---Part I described only two such schemes ---fits best with the known connections on our fibre bundle: in particular, with the Singer-DeWitt connections we have expounded  in the Appendix. \\

Good questions indeed: but sufficient unto the day is the labour thereof . . .

\appendix

\section*{APPENDIX}

\section{The Singer DeWitt connection: construction and examples}
In this Appendix, we will give an introduction of the Singer DeWitt (SdW) connection, as described in Section \ref{subsec:42connn}. 

We start by giving an explicit formula for both the connection and its associated curvature. Then we apply this definition in different theories: in Section \ref{sec:SdW_particles},  to Newtonian point-particles, with translation and rotational symmetry; in Section \ref{sec:SdW_YM}, to pure Yang-Mills theory and its standard gauge symmetry; and in Section \ref{sec:SdW_GR}, to the space of Riemannian metrics, with diffeomorphism symmetry.

As explained in \cite[Sec. 4]{GomesHopfRiello}: given a $\G$-invariant metric on field space, i.e. $\bb G(\bullet, \bullet)$ such that $\bb{L}_{\xi^\#}\bb G=0$ for all $\xi\in \fG$, we will find a unique connection $\varpi$ associated to it (see Equation 4.5 (ibid).\footnote{ Indeed, the metric need not even be completely preserved by the group action: it suffices that it preserves the orthogonal space to the orbits (see Equation 4.6 (ibid)).}
 
  In more detail: A configuration-space metric and a vertical direction supply enough ingredients to define a connection if and only if the directions orthogonal to the orbit remain orthogonal by pull-back along the group action. This is particularly straightforward if the orbits are Killing directions; i.e., the fundamental vector fields $\xi^\#$ are Killing   fields of $\bb G$, i.e.
    \begin{align}
    \fLie_{\xi^\#} \bb G = 0
    \qquad \text{for all} \;
    \xi \in \text{Lie}(\G)\,.
    \label{eq_GG_Killing}
    \end{align}
 For such a field-space metric $\bb G$, we \emph{define} the \emph{Singer-DeWitt (SdW) connection} 
 $\varpi$ by demanding the following orthogonality relation:   
  \begin{equation}
  \bb G(\xi^\#, \hat H(\bb X)) \equiv \bb G(\xi^\#, \bb X - \varpi(\bb X)^\#) = 0,
  \label{eq4.3}
  \end{equation}
for all $\xi \in \text{Lie}(\G)$ and all $\bb X \in \mathfrak{X}^1(\F)$. The preceding equation holds pointwise on  $\F$, where the vector field $\bb X$ identifies, for each configuration $\varphi\in\F$, a tangent vector $\bb X_\varphi \in {\rm T}_\varphi \F$. But we will omit subscripts. Here, $\hat H$ stands for the horizontal projection induced by $\bb G$, so we cold write $\varpi(\cdot)^\#=\hat V(\cdot)$.

Formally,  equation \eqref{eq4.3}, defining an SdW connection, can be solved for $\varpi$ as follows. Let $\bb Q_{ab}$ be the pullback to $\fG$ under $\cdot^\#$ of the metric induced from $\bb G$ on the fibres as expressed in the $\{\tau_a\}$ basis of the Lie-algebra:
  \be
  \bb Q_{ab}=\bb G(\tau_a^\#, \tau_b^\#),
  \label{eq_Qab}
  \ee
and $\bb Q^{ab}$ its inverse. Note that $\bb Q_{ab}$ does not in general coincide with the (point-wise extensions of the) Killing form in $\fg$ (the natural inner product of the Lie-algebra).

  Expanding $\varpi = \varpi^a \tau_a$, equation \ref{eq4.3} can be written as $\bb G(\tau_a^\#, \bb X) = \bb Q_{ab} \fI_{\bb X} \varpi^b$, which is readily inverted as
  \begin{align}\label{eq:varpi_abstract_solution}
  \varpi = \bb Q^{ab} \bb G (\tau_b^\#, \cdot) \tau_a\,.
  \end{align}
   Note that $\bb G(\xi^\#, \cdot)$ accepts field-space vectors and hence defines a one-form in field-space.

From the last equation, we immediately obtain the first fundamental property required of a connection-form: $ \varpi(\xi^\#) = \xi$. See \cite[Sec. 4.1, equations 4.7 and 4.8]{GomesHopfRiello} for a proof that an $\varpi$ defined by the procedure above will also transform correctly under gauge transformations if $\xi^\#$ is a Killing vector of $\bb G$. 

 The resulting relationship between a field-space-metric $\bb G$ and the curvature $\fF$ of the associated $\varpi$ is: 
  \begin{equation}\label{eq:curvature_and_metric}
  \bb G \big( \bb F(\bb X, \bb Y)^\#, \xi^\#\big) = \dd (\bb G (\xi^\#)) (\hat H (\bb Y), \hat H (\bb X))
  \quad \text{for all} \;
  \xi \in \text{Lie}(\G), ~\dd \xi = 0,
  \end{equation}
  and any $\bb X, \bb Y \in \mathfrak{X}^1(\Phi)$. On the right hand side, $\bb G(\xi^\#)\equiv\bb G(\xi^\#,\cdot)$ is a one-form on field-space, so $\dd \bb G (\xi^\#)$ is a two-form. By horizontally projecting the dummy vector fields $\bb X, \bb Y$ on the right hand side, we are taking the horizontal-horizontal part of that two-form. Formally solving for $\bb F$, we get
  \begin{align}
  \bb F = \bb Q^{ab} \big(\dd \bb G(\tau_b^\#) \big)_{HH} \tau_a\,,\label{eq:curv_formula}
  \end{align}
  which is \emph{the main result of this section.} Note that, in these formulas, $\dd$ acts on the one-form $\bb G(\xi^\#)$. Even if $\xi$ is taken to be configuration-independent; i.e., $\dd \xi=0$, the operator $\cdot^\#$ generically introduces configuration-dependence.  For a proof of \eqref{eq:curvature_and_metric} (and thus the origin of \eqref{eq:curv_formula}) see \cite[Sec. 4.2, equation 4.12]{GomesHopfRiello}.

\subsection{The SdW connection for Newtonian particles}\label{sec:SdW_particles}
Given a  Singer-DeWitt connection, $\varpi$, the best-matched configurational velocities will be given by just the horizontal projection with respect to $\varpi$. 

 To explicitly obtain the Singer-DeWitt connection in  the case of $N$ Newtonian particles and compare it with best-matching, we will take the Lagrangian to be of the form 
 \be\label{eq:Lagrange}\mathcal{L}=K(\rr, \dot\rr)-V(\rr) ;
 \ee
  where, in this non-relativistic configuration space of $N$ particles, we revert to the standard dot notation; i.e. $\dot \rr_\alpha$. The total kinetic energy of the system, which will determine our configuration space metric, is 
 \be\label{eq:kinetic}K(\dot\rr)=\frac12\sum_\alpha m_\alpha |\dot\rr_\alpha|^2=\frac12\sum_\alpha m_\alpha (\dot\rr_\alpha\cdot \dot \rr_\alpha)\,, \ee
 where $\cdot$ is the Euclidean inner product in $\R^3$. The kinetic term $K$ is equivalent to a choice of inner product:
 \be\label{eq:config_ip}\bb G(\dot\rr, \dot\rr')= \frac12\sum_\alpha m_\alpha \dot\rr_\alpha\cdot \dot \rr'_\alpha.\ee

 Equation \eqref{eq:config_ip} is clearly invariant under the time-independent transformations: 
 \begin{subequations}
 \begin{align}
  \rr_\alpha&\mapsto \rr_\alpha+\mathbf{v}\\
   \rr_\alpha&\mapsto \rr_\alpha R \,.
\end{align} \end{subequations}
Infinitesimally, the group actions of $\bb R^3$ and $\SO(3)$ correspond to, respectively (for $\mathbf v\in \R^3$ and $\xi\in \mathfrak{so}(3)$):
\begin{subequations}\label{eq:gts}
 \begin{align}
  \delta_{\mathbf v}\rr_\alpha &=\mathbf{v}\\
   \delta_{\xi}\rr_\alpha &=\rr_\alpha \xi\,.
\end{align} \end{subequations}
In terms of the Lie derivative (where $\delta_\xi\rr=:\xi^\#$),
$$\bb L_{\xi^\#}\bb G(\bb v, \bb v' )=\frac{1}{t}\lim_{t\rightarrow 0}(\bb G(\bb v, \bb v')- \bb G(\bb v \xi, \bb v' \xi))=0\,.
$$
In other words: the metric has Killing directions along $\xi^\#$.

For translations, the generators of the algebra $\bb R^3$ can be taken to be the unit vectors, $\tau_a=e_a$, given by $\{e_x, e_ y, e_ z\}.$ The vertical metric $\bb Q^{ab}=\delta^{ab}$, and therefore 
$$\varpi_{\text{\tiny trans}}(\dot \rr)= (\sum_\alpha m_\alpha \dot \rr_\alpha \cdot e_a) e_a= \sum_\alpha m_\alpha \dot \rr_\alpha, 
$$ since $\sum |e_a\rangle\langle e_a|$ is the identity operator. 

An associated connection-form defines, infinitesimally along the trajectory of the system, the standard of linear translations. The translational connection therefore yields the linear momentum of the configurational velocity. Horizontal motion  coincides with a choice of coordinate system for which the total \textit{linear} momentum vanishes. In other words, for an arbitrary velocity $\dot \rr_\alpha$, the horizontal, or best-matched velocity $\dot\rr_\alpha-\varpi_{\text{\tiny trans}}(\dot \rr)$ has vanishing linear momentum. From \eqref{eq:curv_formula}, the curvature clearly vanishes, since neither $\bb G$ nor $\tau_a^\#$ depend on the configuration. This is the main difference compared with the rotational case, which we now assess.

For rotations, we get a vertical metric that is just the moment of inertia tensor:
 \be\label{eq:MIT}
 \bb Q_{ab}=M_{ab}=\frac12\sum_\alpha m_\alpha\left(|\cc_\alpha|^2\delta_{ab}-c_{\alpha a} c_{\alpha b}\right)\,.
 \ee
Because 
  \be\label{eq:intermediary}\bb G(\cdot, J_b^\#)= \frac12\sum_\alpha m_\alpha \dd c^i_\alpha  c^k_\alpha \epsilon_{ibk}, 
\ee
where $\dd c^i_\alpha$ is a basic configuration space 1-form (like $\d x$ would be in space-time), we obtain the following expression for the connection-form from \eqref{eq:varpi_abstract_solution}:
 \be\label{eq:varpi_rot}
 \varpi(\dot \cc)=(M^{-1})^{ab} \left(\frac12\sum_\alpha m_\alpha \epsilon_{bij} c_\alpha^i \dot c_\alpha ^j\right ) J_{(a)}\,.
 \ee
 Given an infinitesimal change of configuration, \eqref{eq:varpi_rot} provides the necessary rotation for that change to carry no angular momentum. By the properties of the connection-form---arising from orthogonality to the fibre with respect to a $\G$-invariant metric---this adjustment is gauge-covariant; i.e., it does not depend on the orientation of the configuration that we started from. The connection-form defines a standard of orientation infinitesimally along a curve.
 
It is also easy to  write the vertical projection. From \eqref{eq:varpi_abstract_solution}, we have $\hat V=\bb Q^{ab} \bb G (J_b^\#, \cdot) J^\#_a$.\footnote{By writing the basis $J^\#_a=|\ell_a\rangle$ and $\bb G (J_b^\#, \cdot) =\langle \ell_b|$ to match \citet{Littlejohn1997}'s notation, we obtain their equation 5.42: $\Pi_V= |\ell_a\rangle (M^{-1})^{ab} \langle \ell_b|$.} If we use equation~\eqref{eq:MIT} and \eqref{eq:varpi_rot}  we obtain: 
\be\label{eq:vert_proj} \hat V(\dot \cc)=\varpi(\dot\cc)^\#=\sum_\alpha m_\alpha \dot\cc_\alpha\times \cc_\alpha=\mathbf{L}\,,
\ee
which is the angular momentum. The horizontal projection is just its complement: $\hat H=\bb 1-\hat{V}$. Given a generic centre-of-mass configurational velocity $\dot \cc_\alpha$, the best-matched, or horizontal velocity $\dot \cc_\alpha-\hat V(\dot \cc)$ has vanishing angular momentum. 

 Curvature implies that, for a closed loop in the base space, the orientation may change even for motion with zero angular momentum. We can now write the curvature using \eqref{eq:curv_formula}. First, from \eqref{eq:intermediary}:
$$\dd\bb G(\cdot, J_b^\#)= \frac12\sum_\alpha m_\alpha \epsilon_{ibk}\,\dd c^i_\alpha\curlywedge \dd c^k_\alpha, 
$$ where $\curlywedge$ is the exterior differential for forms in configuration space. From this equation it is apparent that if the group action on configuration space did not depend on the configuration, the exterior derivative $\dd$ would have nothing to act non-trivially on. We would then obtain $\dd\bb G(\cdot, J_b^\#)=0$, and vanishing curvature as a consequence. This is what occurs for translations. As it stands, the curvature for the rotational bundle can be written as:
\be \bb F(\dot c, \dot c')=(M^{-1})^{ab} \left(\frac12\sum_\alpha m_\alpha \epsilon_{bij} \hat{H}(\dot \cc_\alpha)^i \hat H(\dot \cc'_\alpha )^j\right ) J_{(a)}\label{eq:curv_rot}\,,
\ee
which only depends on the base space through the moment of inertia tensor and the horizontal projections.

 \subsection{The SdW connection for Yang-Mills}\label{sec:SdW_YM}
The prime example for an ultralocal field-space metric is the gauge supermetric for Yang-Mills (YM) theories.
	In the following, to emphasize our neglecting matter fields, we introduce the notation $\F_{\rm pYM}$ to indicate the field-space of `pure' Yang--Mills theory. This is constituted by:
	 $\fg$-valued 1-forms over the spacetime manifold\footnotemark~$M$,%
	\footnotetext{  One could also take the (less popular) parametrization of the degrees of freedom for YM  given by $\omega$, a connection on a principal fiber bundle $P$, with the base space being spacetime $M$ and with fibers isomorphic to $G$. Then the relation between $A$ and $\omega$ requires a section $\sigma:M\rightarrow P$, and is given by $A=\sigma^*\omega$. The distinction between these parametrizations is important for non-trivial bundles, where $\omega$ exists globally but $A$ exists only locally.  Although we will stick to the more popular ``physicists' " parametrization, i.e. $A$,  our formalism can be readily adapted with minimal modifications to treating $\omega$ as the fundamental variable. \label{ftnt:PFB}}
	\be
	A = A^a_\mu(x) \tau_a\d x^\mu \in \Lambda^1(M, \fg),
	\ee
	where $\fg= \text{Lie}(G)$ and $\{ \tau_a \}_a$ is an orthogonal basis of the latter. The fundamental vector fields are:
	\be
	\xi^\# = \int \delta_\xi A \frac{\dd}{\dd A},
	\label{xi_hash}\ee
where 
\be
	\delta_\xi A = \D \xi := \d \xi + [A,\xi]
	\label{eq_infin_gtransf}
	\ee
	with $[\cdot,\cdot] $ the Lie bracket on $\fg = {\rm Lie}(G)$, extended pointwise on $M$ to $\fG$.
	
	 In this field-space, our constructions used a positive-definite supermetric. In the Lorentzian case, such an assumption is hard to substantiate. Therefore, we restrict our attention to two cases: (\textit{i}) $M$ is spacetime, but with Euclidean signature; or (\textit{ii}) $M=\Sigma$ represents a (portion) of a spacelike Cauchy hypersurface, in which case: spacetime admits a Lorentzian signature, field-space is understood to be the space of field configurations on $\Sigma$,  and $d$ is the dimension of $\Sigma$ rather than spacetime.

	As a starting point, before considering more complex situations: consider the case where $\Sigma$ is a spacelike {\it compact} Cauchy surface, that is assumed for now to have no boundary and trivial de Rham cohomology. Here $g_{ij}$ is a {\it fixed} positive-definite metric on $\Sigma$; (it is a background structure, not part of field-space). Notice that $\bb G^{\rm g}$ is independent of $A$.

	The gauge supermetric contracts variations of the gauge field, $\bb X = \int \bb X \frac{\dd}{\dd A}\in \mathrm T_A \F_{\rm pYM}$, as in%
	\footnote{For $\bb G^{\rm g}$ to be dimensionless (in units of $\hbar$), it has to be multiplied by $e^{-2}$, where $e$ is the Yang--Mills coupling constant.}
	\begin{equation}\label{metric_YM}
	\bb G^{\rm g} (\bb X, \bb Y)= \int_\Sigma \d^dx\, \sqrt g g^{ij}  \delta_{ab} \bb X_i^a \bb Y_j^b ,
	\qquad 
	{\bb X}, \bb Y \in {\rm T}_A \F_{\rm pYM}.
	\end{equation}

	To illustrate the general features of the SdW connections, let us solve for the connection arising for the  spatial  supermetric  just introduced. We have
	\begin{align}
	0 ={}
	\bb G^{\rm g} (\xi^\sharp, \bb X -   \varpi({\bb X})^{\ \#})
	={}& 
	\int \d^d x\,\sqrt g g^{ij} \delta_{ab} \D_i \xi^a\big(\bb X_j^b - \D_j (\varpi({\bb X})^{b})\big) \nonumber\\
	={}& - \int \d^d x \, \sqrt{g} \delta_{ab}\xi^a \big( \D^i \bb X_{i}^b - \D^i \D_i (\varpi({\bb X})^{b})\big) ,
	\label{explicit_hor}
	\end{align}
	and, using the arbitrariness of $\xi^a(x)$ and $\bb X^a_i(x)$, we  deduce 
	\be
	{\quad\phantom{\Big|}
		\D^2\varpi =  \D^i \dd A_{i},
		\label{varpi_YM}
		\quad}
	\ee
	where $\D^2 := (\D^i \D_i)$ is the gauge-covariant Laplacian. The horizontal vector fields are the kernel of $\varpi$. Contraction with \eqref{varpi_YM} shows that, in this simple case, the horizontal vector fields are  those which are (covariant-)divergence free: $\D^i\bb X_i = 0$, since by definition $ \dd A_i({\bb X}) = \bb X_i$ for any vector $\bb X$.
	We see that SdW connections are generically of the form: `inverse Laplacian of divergence'.

	In an Abelian pure Yang-Mills theory, $\D = \d$, and the above   equation  \eqref{varpi_YM} becomes a Poisson equation for field-space one-forms. The equation therefore has a unique solution (up to a constant).
	For non-Abelian theories the relevant Laplacian operator is field-dependent (or background-dependent) and the defining equation for $\varpi$ becomes more involved.\footnote{ For a space topology of $S^3$,  gauge group $SU(2)$, and appropriate analytic conditions on the Yang-Mills connection, the kernel of the Laplacian is the  stabilizer of $A$, i.e., Lie algebra elements $\xi$ with $D \xi = 0$. Most Yang-Mills connections have a trivial stabilizer, and the Laplacian is invertible on those.}
	
	 To compute the curvature of $\varpi$, it is most convenient to use equation \eqref{eq:curvature_and_metric}, rather than trying to compute it directly from $\fF = \dd \varpi + \tfrac12 [\varpi,\varpi]$.
	Consider a field-space constant $\xi$, i.e. $\dd \xi = 0$. We have $\D_i \xi^a = \partial_i \xi^a+ f^a{}_{bc} A_i^b\xi^c$ with $f^a{}_{bc}$ the structure constants of $\fg$.
	Then
	\begin{align}
	\dd \bb G^{\rm g}(\xi^\#) ={}& \dd \int \d^d x \sqrt g g^{ij} \delta_{ab} \D_i \xi^a \dd A_j^b
	= - \int \d^d x \sqrt g g^{ij} \xi^c f_{abc} \dd A^a_i \dd A^b_j \qquad (\dd \xi = 0).
	\end{align}
	The horizontal projectors on the right hand side of equation \eqref{eq:curvature_and_metric} have the effect of replacing $\dd$ with $\dd_H$ in the last line. On its left hand side, after an integration by parts, we have
	\be
	\bb G^{\rm g}(\fF^\#, \xi^\#) = - \int \d^d x \sqrt g \xi^a \delta_{ab} \D^i \D_i \fF^b.
	\ee
	Hence, by equating the two and using the cyclicity of the structure constants (i.e. for a compact semisimple Lie algebra), as well as the arbitrariness of $\xi$, we obtain 
	\be
			\D^2 \fF =  g^{ij} [\dd_H A_i ,\dd_H A_j ],
	\label{Singercurvature}
	\ee
	or more explicitly, $\D^2 \fF^a =  f^a{}_{bc}  g^{ij} \dd_H A^b_i \dd_H A^c_j $.
	This result for the curvature of the Yang-Mills DeWitt connection was r  first given  by Singer \cite{Singer:1978dk}, in a context where $\Sigma$ is an Euclidean spacetime without boundary (rather than a time slice). 
	
	\subsection{The SdW connection for Riemannian metrics}\label{sec:SdW_GR}
	
	We can now be brief. First, we establish the horizontality condition, for a closed manifold, for the space of Riemannian metrics, $g\in$ Riem(M). Here $\G=$ Diff(M); and $\fG=\mathfrak{X}(M)$, i.e. the smooth vector fields over $M$.

Unlike the other two cases discussed in this Section,  the kinetic term of the general relativistic  action in the 3+1 formalism does not give a \emph{bona-fide}, i.e. positive-definite supermetric. The problem is well-known  (\cite{Kuchar_Time}),  and arises  from the subtraction of the product of the trace of the tangent vectors on Riem (see footnote \ref{ftnt:dewitt}). If one chooses a  gauge-fixing of the Hamiltonian constraint so that the volume element $\sqrt{g}\d^3 x$ is constant (see \cite{Gielen2018} for details), then, since the resulting lapse is positive and `pure-trace' directions don't preserve the gauge-fixing condition, we can find a  connection for the remaining spatial diffeomorphisms (see \cite{gomes_riem}) via  the same procedure as we now follow. 

In this Section, we will illustrate the connection with a simplified, non-dynamical, choice of supermetric:\footnote{The DeWitt supermetric with unit lapse would be:  $ \bb G( \bb X, \bb Y)_g:=\int \sqrt{g}\,( g^{ac}g^{bd}-\frac12  g^{ab}g^{cd}) \bb X_{ab} \bb Y_{cd} $. \label{ftnt:dewitt}
} 
\be\label{eq:supermetric} \bb G( \bb X, \bb Y)_g:=\int \sqrt{g}\, g^{ac}g^{bd} \bb X_{ab} \bb Y_{cd}. 
\ee
A vertical element at $g_{ab}$ is, for $X\in \mathfrak{X}(M)$,  of the form $X^\#:=\bb X_{ab}=\nabla_{(a}X_{b)}$. Thus we obtain, as in \eqref{explicit_hor}, by a simple integration by parts, that the horizontality condition is:
\be\label{eq:horizontal}  \bb X\,\,\text{is horizontal iff}\quad \nabla^a \bb X_{ab}=0.
\ee
To obtain the connection, we have to solve 
\be\label{eq:inv_pde}\nabla^a(\bb X_{ab}-\nabla_{(a}\varpi(\bb X)_{b)})=0
\ee
for $\varpi$ in terms of $\bb X_{ab}$. Writing $D$ for the linear differential operator defined by:
\be \nabla^a\nabla_{(a}Y_{b)})=:D(Y)_b,
\ee
for $Y\in C^\infty(TM)$, 
we have  that $D$ is elliptic. It has a non-trivial finite-dimensional kernel when the metric has Killing directions (which we are ignoring). And thus we can invert \eqref{eq:inv_pde}, obtaining:
\be  \varpi_b(\bb X)=D^{-1}(\nabla^a \bb X_{ab}).
\ee
This can be seen as the generator of the vertical transformation that takes $\bb X$ to a horizontal vector, i.e. the generator of a horizontal projector. 

In other words, $\varpi$ takes an infinitesimally different metric and gives back an infinitesimal diffeomorphism. The interpretation of this infinitesimal diffeomorphism is like that of best-matching. It  drags the field-values in one copy of $M$ by whichever infinitesimal diffeomorphism  makes the dragged-along values over the first copy of $M$ be, collectively, as close as possible to the values at the points in the other copy of $M$ to which they have been dragged.  Here of course, the phrase `as close as possible' is to be understood according to $\bb G$.

\subsection*{Acknowledgements}
For comments on this material, we are grateful to: the audience, and especially the organizers, at the DICE conference in Castglioncello; the audience and organizers at the August 2022 QISS meeting;   the participants at the LSE-Cambridge Philosophy of Physics Bootcamp discussion group; seminar audiences in Bonn, Bristol, Oxford and Warsaw;   and Sam Fletcher for comments on a previous version.\\


\end{document}